\documentclass[fleqn,usenatbib]{mnras}

\usepackage{newtxtext,newtxmath}

\usepackage[T1]{fontenc}
\usepackage{ae,aecompl}

\usepackage{graphicx}	
\usepackage{amsmath}	
\usepackage{amssymb}	

\newcommand{\tk}{2010~TK$_7$}

\title[Search for L5 ETs with DECam]{Search for L5 Earth Trojans with DECam}

\author[L. Markwardt et al.]{Larissa Markwardt,$^{1}$\thanks{E-mail: lmmarkwa@umich.edu}
D. W. Gerdes,$^{1,2}$
R. Malhotra,$^{3}$
J. C. Becker,$^{4,1}$
S. J. Hamilton,$^{2}$
\newauthor F. C. Adams$^{1,2}$
\\
$^{1}$Department of Astronomy, University of Michigan, 1085 South University Avenue, Ann Arbor, MI 48109, USA\\
$^{2}$Department of Physics, University of Michigan, 450 Church Street, Ann Arbor, MI 48109, USA\\
$^{3}$Lunar and Planetary Laboratory, The University of Arizona, 1629 E University Boulevard, Tucson, AZ 85721, USA\\
$^{4}$Division of Geological and Planetary Sciences, California Institute of Technology, 1200 E California Blvd, Pasadena CA 91125, USA
}

\date{Accepted XXX. Received YYY; in original form ZZZ}

\pubyear{2020}

\begin{document}
\label{firstpage}
\pagerange{\pageref{firstpage}--\pageref{lastpage}}
\maketitle

\begin{abstract}
Most of the major planets in the Solar System support populations of co-orbiting bodies, known as Trojans, at their L4 and L5 Lagrange points. In contrast, Earth has only one known co-orbiting companion. This paper presents the results from a search for Earth Trojans using the DECam instrument on the Blanco Telescope at CTIO. This search found no additional Trojans in spite of greater coverage compared to previous surveys of the L5 point. Therefore, the main result of this work is to place the most stringent constraints to date on the population of Earth Trojans. These constraints depend on assumptions regarding the underlying population properties, especially the slope of the magnitude distribution (which in turn depends on the size and albedo distributions of the objects). For standard assumptions, we calculate upper limits to a 90\% confidence limit on the L5 population of $N_{ET}<1$ for magnitude $H<15.5$, $N_{ET}=60-85$ for $H<19.7$, and $N_{ET}\ $= 97 for $H=20.4$. This latter magnitude limit corresponds to Trojans $\sim$300 m in size for albedo $0.15$. At H=19.7, these upper limits are consistent with previous L4 Earth Trojan constraints and significantly improve L5 constraints.
\end{abstract}




\section{Introduction}



The population of Earth Trojans (ETs) represents an important (if perhaps not abundant) part of our Solar System because it can provide constraints on the Solar System's dynamical history. This paper reports the results from a new observational survey that searches for ETs near the L5 Lagrange point of the Earth-Sun system. The survey uses the DECam instrument on the Blanco 4 meter telescope. Although no new objects have been detected, this observational program provides the strongest constraints to date on the population of ETs.

\subsection{Significance of Earth Trojans}

Despite their relative proximity to Earth compared to other astronomical objects, the Earth Trojan population is poorly understood. ETs are asteroids that orbit near the Earth-Sun L4 or L5 Lagrange points (60\textdegree\  ahead or behind the Earth, respectively). Trojan asteroids have been discovered accompanying other planets, including Venus, Mars, Jupiter, Uranus, and Neptune \citep{VTs, MTs, JTMorbi, JT_Yoshida_2017, NT, NT_ours, ut}. In contrast, only one ET has been found, \tk, which was discovered serendipitously by WISE \citep{Connors2011}. 
\par Characterization of this population is important as ETs can give a unique insight into our Solar System. Numerical simulations have shown that ETs can have orbits near the L4 or L5 points that are stable on time scales of order the current age of the Earth \citep{Cuk2012, Malhotra2011, Tabachnik2000, Marzari2013}. As a result, the objects in this population could be undisturbed remnants from our primordial protoplanetary disk. In principle, the distribution of ETs can thus be used to constrain Solar System formation models. The chemical composition of these bodies could also provide insight into the properties of the primordial material from which our Solar System formed. 
\par ETs have also been studied as potential lunar impactors. In particular, there appears to be an asymmetry of rayed lunar craters on the leading and trailing sides of the moon \citep{Morota2003}. Both \citet{Gallant2009} and \citet{Ito2010} used numerical simulations to assess whether this asymmetry could be due to impacts of Near Earth Asteroids (NEAs). \citet{Gallant2009} concluded that, within the uncertainties of their simulations and of the observational data, the known population of NEAs were sufficient to account for the observed crater asymmetry, whereas \citet{Ito2010} found that the known NEA population could only account for $\sim50$\% of the asymmetry. An unobserved population of NEAs with very small velocities relative to the moon, like Earth co-orbitals, has been suggested and studied as a solution to this discrepancy \citep{Ito2010, JeongAhn2010}. 

\par However, due to the lack of observational constraints on the ET population, the viability of ETs as this missing impactor population has been investigated only through theoretical models of the population, which may not represent the true distribution of co-orbitals. Detections of the ET populations in both the L4 and L5 Lagrange points --- or robust upper limits --- will provide important constraints on planet formation theories, which consider both the dynamical evolution and the physical properties of rocky Solar System objects. 

\subsection{An Unstable Earth Trojan: \tk}

\par ETs are a particularly difficult population to observe; the L4 and L5 points are always at low solar elongations as observed from the Earth ($\sim$60\textdegree\ ahead or behind the Sun respectively), meaning they can only be observed around dawn or dusk when the sky is still relatively bright. Moreover, due to the geometries of their orbits, ETs are never observed at opposition, and their larger phase angles make them even fainter and harder to detect.

\par These challenges may be part of the reason that only one ET, \tk,  has been observed to date \citep{Connors2011}.
However, the orbit of this object is not consistent with primordial ETs, which are expected to be long-term stable librators near L4 or L5.  
In particular, \tk\ has a large amplitude tadpole orbit, librating between Earth and L3 (the Lagrange point behind the Sun), rather than remaining near L4 \citep{Connors2011, Dvorak2012}. We consider such orbits distinct from primordial ETs, though it is possible for these orbits to also be stable on the order of $\sim$Myrs \citep{Marzari2013, Dvorak2012}. 
\par Numerical integrations indicate that \tk\ has a highly chaotic and short-lived orbit. Estimated lifetimes vary from $\sim$7000 years \citep{Connors2011} to about $\sim$0.25 Myrs  \citep{Dvorak2012}. Taking into account the Yarkovsky effect, \cite{Zhou2018} also concluded that \tk\ is too small to maintain a long-term stable orbit. As a result, this object is most likely an asteroid that was temporarily captured as an Earth co-orbital, rather than a primordial ET. 

\subsection{Previous ET Searches}

\par There have been a few dedicated searches for  ETs. The most recent searches were a ground-based survey for L5 ETs \citep[W\&T98]{Whiteley1998}, an OSIRIS-REx search as the spacecraft flew past the L4 point on its way to the asteroid Bennu \citep[O-R18]{Cambioni2018}, and a search made by the Hayabusa2 spacecraft as it flew past the L5 point on its way to Ryugu \citep{hayabusa}. None of these surveys found any other Earth Trojans. 

\par An upper limit calculation from the Hayabusa results has not yet been published \citep{hayabusa}. \citet{Whiteley1998} placed an upper limit on the ET population of $\sim$3 objects per square degree to R=22.8. \citet{Cambioni2018} placed on upper limit on the population of 73 $\pm$ 22 objects $\sim$210 m (S-type asteroids) to $\sim$470 m (C-type asteroids) in size. For comparison, they also applied their method to the \citet{Whiteley1998} survey and found an upper limit of 194 $\pm$ 116 objects at the OSIRIS-REx limiting magnitude \citep{Cambioni2018}. These limits imply that there could still be tens to hundreds of undiscovered ETs. Moreover, the \citet{Whiteley1998} search at L5 only covered a 0.35 sq. deg. area near L5. From these relatively large upper limits and limited survey coverage, it is clear that the ET population is far from being completely, or even sufficiently, characterized.

\vspace{5mm}
In this paper we present the results of a new survey directed at the Sun-Earth L5 point with DECam on the Blanco 4m telescope. In Section \ref{our_survey}, we provide a description of the observations that make up our survey. In Section \ref{pipeline}, we provide a description of the pipeline to reduce the data, extract moving objects, and create a catalog of ET candidates. In Section \ref{objects}, we present all of the moving objects found by our survey. In Section \ref{limits}, we present the upper limits on the ET population derived from the results of our survey. In Section \ref{discussion}, we discuss our findings and their implications.

\begin{figure*}
\includegraphics[width=1\textwidth]{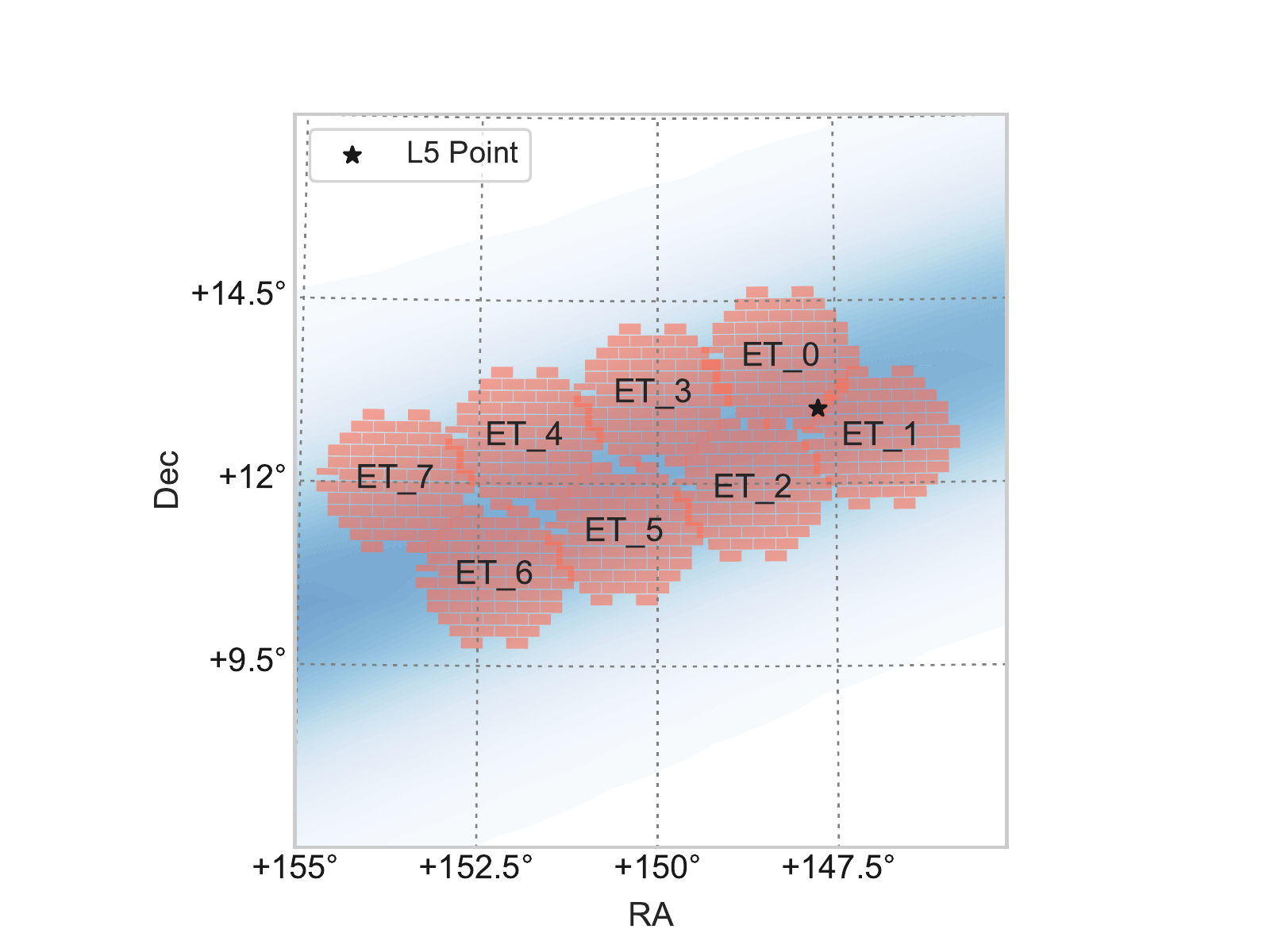}
  \caption[t]{The area covered by our survey. The star marks the location of the L5 point. The relative density of synthetic ET objects (see Sec.~\ref{fakes}) is shown in the background in blue. Each rectangle corresponds to a chip in one of the 8 DECam fields. In total, these fields cover $\sim$ 24 sq.~deg.~near L5.
  }
  \label{fig:point}
\end{figure*}

\section{DECam Survey} \label{our_survey}

\par Our survey utilized the VR filter on the Dark Energy Camera \citep[DECam,][]{Flaugher_2015}, which is a prime focus  imager  on  the  4-meter  Blanco telescope at the Cerro Tololo Inter-American Observatory. The VR filter is the widest filter on DECam, making it easier to detect faint ETs. The sky coverage for this survey is shown in Fig.~\ref{fig:point}. We observed shortly after sunset on the night of June 16, 2018, covering an area near Earth's L5.  

DECam has 61 active CCD chips that cover 3  sq.~deg.~on the sky. As shown in Fig.~\ref{fig:point}, our survey consisted of 8 fields, covering an area of 24 sq. deg. ($\sim$ 70 times larger than Whiteley \& Tholen, 1998). See Table \ref{table:pointings} in Appendix for complete list of exposures. Since stable ETs may have libration amplitudes of up to 30${^\circ}$ but low inclinations, the fields were arranged to proceed eastward from the L5 point within roughly $\pm 2^{\circ}$ of the ecliptic.  

\par Two consecutive 15 second exposure were taken of each field, and this sequence was repeated 4 times at approximately 15-minute intervals. The two short exposures were used to identify objects that were transient over $\sim$ 30 seconds of readout time between exposures. Since ETs are expected to have a rate of motion across the sky of $\sim$ 1 deg/day (see Sec.~\ref{be_ets}), an exposure time of 15 seconds was chosen to avoid trailing of these objects in the individual images, which would make these objects harder to detect in the individual exposures. The subsequent visits were used to track the motion of the object and identify a tracklet for a potential Solar System object. 

\par This sampling strategy did \textit{not} allow us to fit an orbit to detected Solar System objects, because the observational arcs are too short. Therefore, we were not able to dynamically classify any of the objects found in our survey (as was done in \citealp{MTs}). Any new objects found in this survey would require additional followup to be classified as ETs, but they could be identified as ET candidates due to their characteristic rate of motion across the sky of $\sim$1 deg./day (see Sec.~\ref{be_ets}).


\section{Reduction Pipeline} \label{pipeline}

We reduced the survey data and extracted Solar System objects, in particular ETs, by the following steps:\footnote{Code for this paper available at: \url{https://github.com/markwardtla/EarthTrojanSearch-2018}}

\begin{enumerate}
  \item DECam Community Pipeline \citep{Valdes2014}: for basic data reduction
  \item \textit{Source Extractor} \citep{Bertin1996}: for source identification
  \item \textit{Bullseye}: a pipeline we developed for this project to link transient sources
  \item Verification by eye/catalog creation: for removal of false positives
\end{enumerate}

\subsection{Source Extractor} \label{se}

We used Source Extractor (SE) to identify sources in each of the exposures. Table \ref{table:se_table} shows the SE parameters used for this analysis. The DETECT\_THRESH parameter was set to 1$\sigma$; it was set this low for completeness at very faint objects, as many anomalous sources could be identified/filtered out (see Sec. \ref{be_stationary}, \ref{be_lines}).
\par In particular, after running SE, we remove any sources with a center pixel flagged as bad by the DECam Community Pipeline. This pipeline flags bad/compromised pixels (e.g.~known detector defects), saturated pixels, bleed trails, and transient artifacts \citep{Valdes2014}. We do not use the transient flags in order to ensure that we are not losing transient Solar System objects. This step removes hundreds of blatantly bad/compromised sources from each chip.

\begin{table}
\caption{Source Extractor Parameters. The filter called ``all-ground'' is a 3x3 convolution mask with FWHM = 2 pixels.}
\begin{center}
\begin{tabular}{ll}
\hline
\multicolumn{2}{c}{Extraction Parameters}\\
\hline
DETECT\_TYPE&CCD \\
DETECT\_MINAREA&6 \\
DETECT\_THRESH&1.0 \\
ANALYSIS\_THRESH&1.0 \\
FILTER&Y \\
FILTER\_NAME&all-ground \\
DEBLEND\_NTHRESH&32 \\
DEBLEND\_MINCONT&0.005 \\
CLEAN&N \\
\hline
\multicolumn{2}{c}{WEIGHTing Parameters}\\
\hline
WEIGHT\_TYPE&MAP\_VAR\\
\hline
\multicolumn{2}{c}{Background Parameters}\\
\hline
BACK\_TYPE&AUTO \\
BACK\_VALUE&0.0 \\
BACK\_SIZE&256 \\
BACK\_FILTERSIZE&3 \\
\end{tabular}
\label{table:se_table}
\end{center} 
\end{table}

\subsection{Bullseye} \label{be}

Once the SE catalogs for each exposure were created, we linked these sources together to identify transient Solar System objects, particularly ETs. Based on the cadence of our survey, the best possible detection of a Solar System object would be 8 individual detections (one for each exposure) in a nearly straight line, due to the short arc of these observations ($\sim$ 1 hour). We created \textit{Bullseye} to  identify sets of transient detections that fit to a line and have a speed consistent with nearby Solar System objects (see Table~\ref{table:be_table}), and to flag those whose speed is consistent with ETs.  This process is run for each survey field, chip by chip. We do not attempt to link objects that crossed chip boundaries; from our recovery rate of known Solar System objects, we determined that chip-crossing objects are a small fraction of the population and some may still have linkable observations in at least one of the chips. The steps in this pipeline are described in more detail in the following sections.

\begin{table}
\caption{Bullseye Parameters. R is the stationary source radius (measured by SE). See Sec.~\ref{be} for discussion of these parameters.}
\begin{center}
\begin{tabular}{ll}
\hline
\multicolumn{2}{c}{Transient Identification}\\
\hline
Transient metric&\textgreater 0.25"\\
Stationary metric&\textless 0.5"\\
Min. detections for stationary&3\\
Bright Stationary metric&\textless 2"\\
Min. detections for bright stationary&7\\
Transient ring min. radius&R\\
Transient ring max. radius&3.5R\\
Transient ring min. density&0.035 sources/arcsec$^2$\\
\hline
\multicolumn{2}{c}{Linking}\\
\hline
Min. detections for link&5 \\
Min. speed&0.1 deg./day \\
Max. speed&2.0 deg./day \\
Cluster eps&0.05 \\
Min samples for cluster&4 \\
\hline
\multicolumn{2}{c}{Flagging ET Candidates}\\
\hline
Min. speed&0.75 deg./day \\
Max. speed&1.25 deg./day \\
\end{tabular}
\label{table:be_table}
\end{center} 
\end{table}

\subsubsection{Identify Transient Sources} \label{be_transients}

The pipeline starts by identifying sources from the SE catalogs that are transient between exposures. Using astropy's  \texttt{match\_to\_catalog} \citep{astropy1,astropy2}, we compare the catalog of the sources between each of the sets of two exposures. If the distance to the nearest source between the two catalogs is more than the transient metric (i.e.~there is no source at the same position in the next exposure, see Table \ref{table:be_table}), the source is considered to be transient. This comparison must actually be done both ways (i.e.~compare the first catalog to the second and vice versa) because it is possible for a transient source to only show up in the second of the two catalogs due to differences in observing conditions between exposures. Only objects that are transient between the sets of two exposures will continue on to the next step.

\subsubsection{Cut Stationary Sources} \label{be_stationary}

Stationary objects (sources with the same position in a set of two exposures) should not be included in the transient catalog after the previous step. However, stationary sources with inconsistent astrometry or faint stationary sources which were not detected in one of the two exposures in a set could have been erroneously included in the transient catalog and needed to be removed. We use astropy's  \texttt{search\_around\_sky} method \citep{astropy1,astropy2} to compare objects in the transient catalog from the previous step to an SE catalog which includes the source detections in all of the exposures for a field. If a transient object's position matches to too many objects in the full SE catalog within the search radius (see Table \ref{table:be_table}), it will be cut. In other words, this step removes objects which have the same position in 3 out of the 8 exposures. To avoid cutting true transients, we choose parameters for these cuts that are more lenient to avoid removing actual transients from our search. We also removed rings of scattered light around stationary sources that meet the ``bright stationary" criteria in Table \ref{table:be_table}. 

\subsubsection{Connect Transients} \label{be_connect}

After cuts to the transient catalog, we link these detections between exposures together to identify potential Solar System objects. Detections are linked if their motion between exposures is consistent with the motion of a Solar System object. In particular, we assumed a min and max speed typical of nearby Solar System objects, primarily main belt asteroids (see Table \ref{table:be_table}), but did not assume a direction of motion. There are classes of Solar System objects that would not make our speed cut, such as very nearby asteroids or distant, outer Solar System objects. However, neither of these populations are the focus of this search, and the expected ET speed of 1 deg./day is well within our search space. 
\par With a min and max speed but no restriction on direction, the search space for another detection is an annulus. Therefore, for each transient detection (the ``anchor" transient), we used \texttt{search\_around\_sky} to identify other transient objects that fall within that search annulus in each of the subsequent exposures. Anchor transients can come from any of the exposures to account for sources near our limiting magnitude that were not detected in the first exposure and objects near the edge of the chip. At the end of this step, we have a list of transient sources that are linked to subsequent detections that would be consistent with the rate of motion of Solar System objects.

\subsubsection{Fit Lines of Motion} \label{be_lines}

If all of the detections connected to an anchor transient from the previous step were of the same Solar System object, we would expect them to have the same velocity at each observation. Here, we define the velocity as the object's rate of motion across the sky in deg./day (speed) and position angle (direction) as measured with respect to the anchor point.  Therefore, all of these transient detections need to have similar velocities in order to identify them as the same object. 

\par To group these detections by velocity, we used \texttt{DBSCAN} (``Density-Based Spatial Clustering of Applications with Noise") from the \textit{scikit-learn} package.\footnote{\url{https://scikit-learn.org/stable/index.html}} This method is able to identify regions with a high density of points in Cartesian space; specifically, a cluster must have at least some minimum number of points within a certain distance (see ``eps" parameter in  Table \ref{table:be_table}). However, the velocities of our transient detections are in polar coordinates (a speed with an angular direction). Therefore, we converted the speed and position angle into a Cartesian space to then identify detections with clustered velocities. If this cluster contained at least 5 detections from different exposures of the linked object, we add it to our transient object catalog. Therefore, in the end we have a final set of detections that are linked together as a candidate tracklet.

\begin{figure}
\includegraphics[width=0.5\textwidth,keepaspectratio=true]{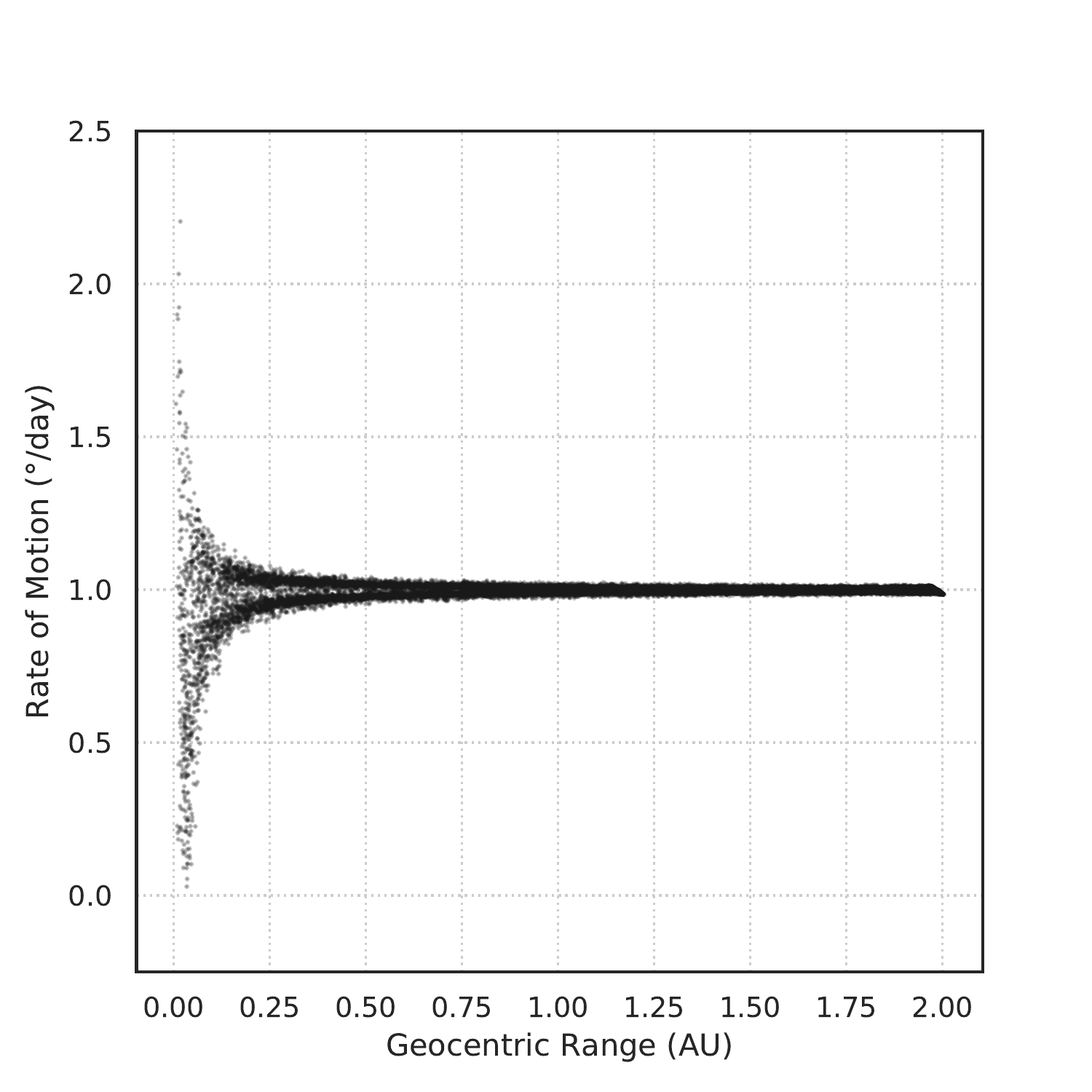}
\includegraphics[width=0.5\textwidth,keepaspectratio=true]{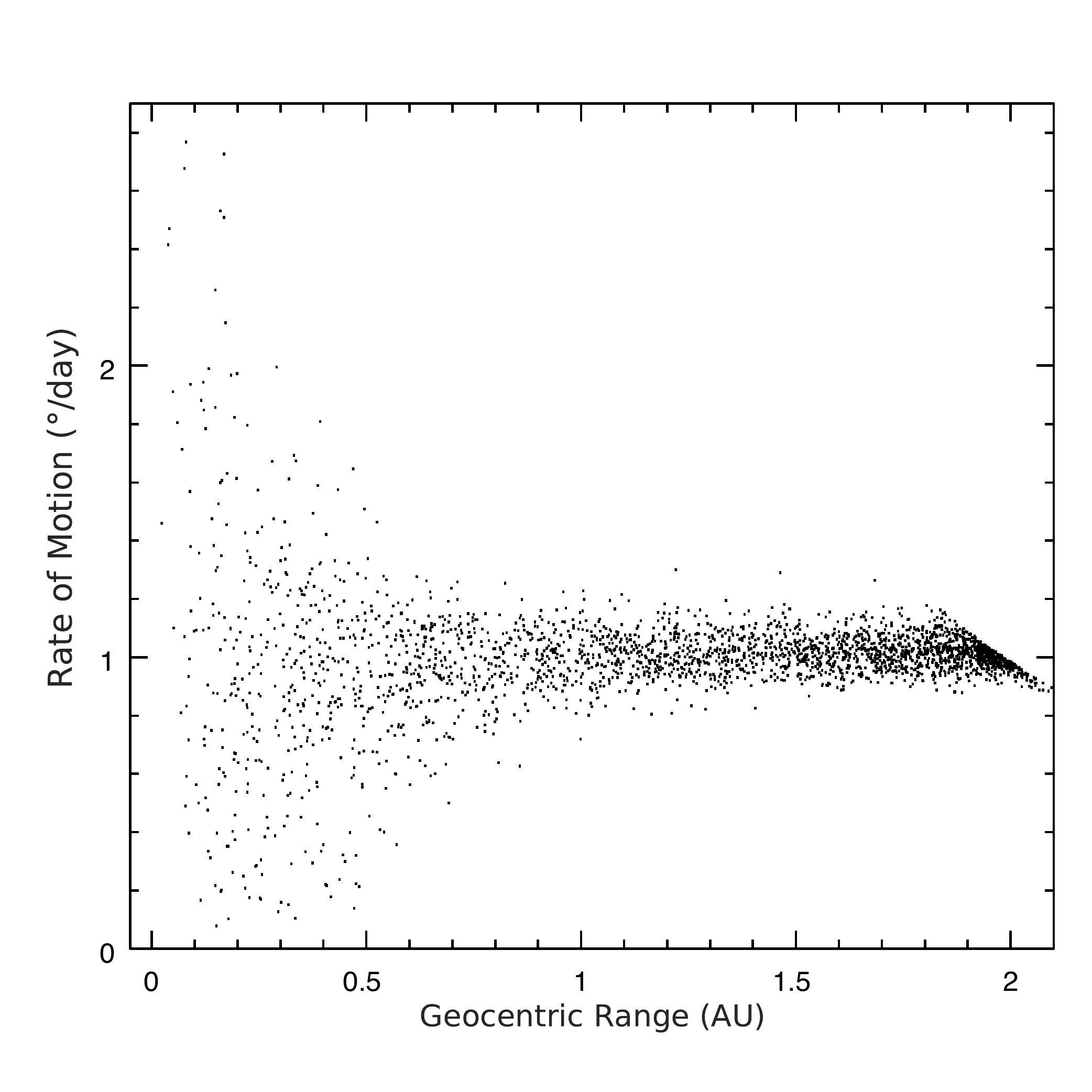}
  \caption[t]{\textit{Top}: The rate of motion on the sky for the ``Strict" set of synthetic ETs (see Sec.~\ref{fakes}). Other than at small geocentric ranges ($\lesssim0.1$ AU), the ETs have a well constrained speed of 1 deg./day. \textit{Bottom}: Same as \textit{Top} but for the ``Jupiter Trojan" set of synthetic ETs (synthetic ETs generated based on observations of Jupiter Trojans; see Sec.~\ref{fakes}). These objects also have a well constrained speed of $\sim$1 deg./day for geocentric ranges \textgreater0.5 AU. There is more dispersion in the speed, but this is expected as this set of ETs was generated based off of observations of Jupiter Trojans. 
  }
  \label{fig:et_speed}
\end{figure}
\begin{figure*}
\includegraphics[width=\textwidth,keepaspectratio=true]{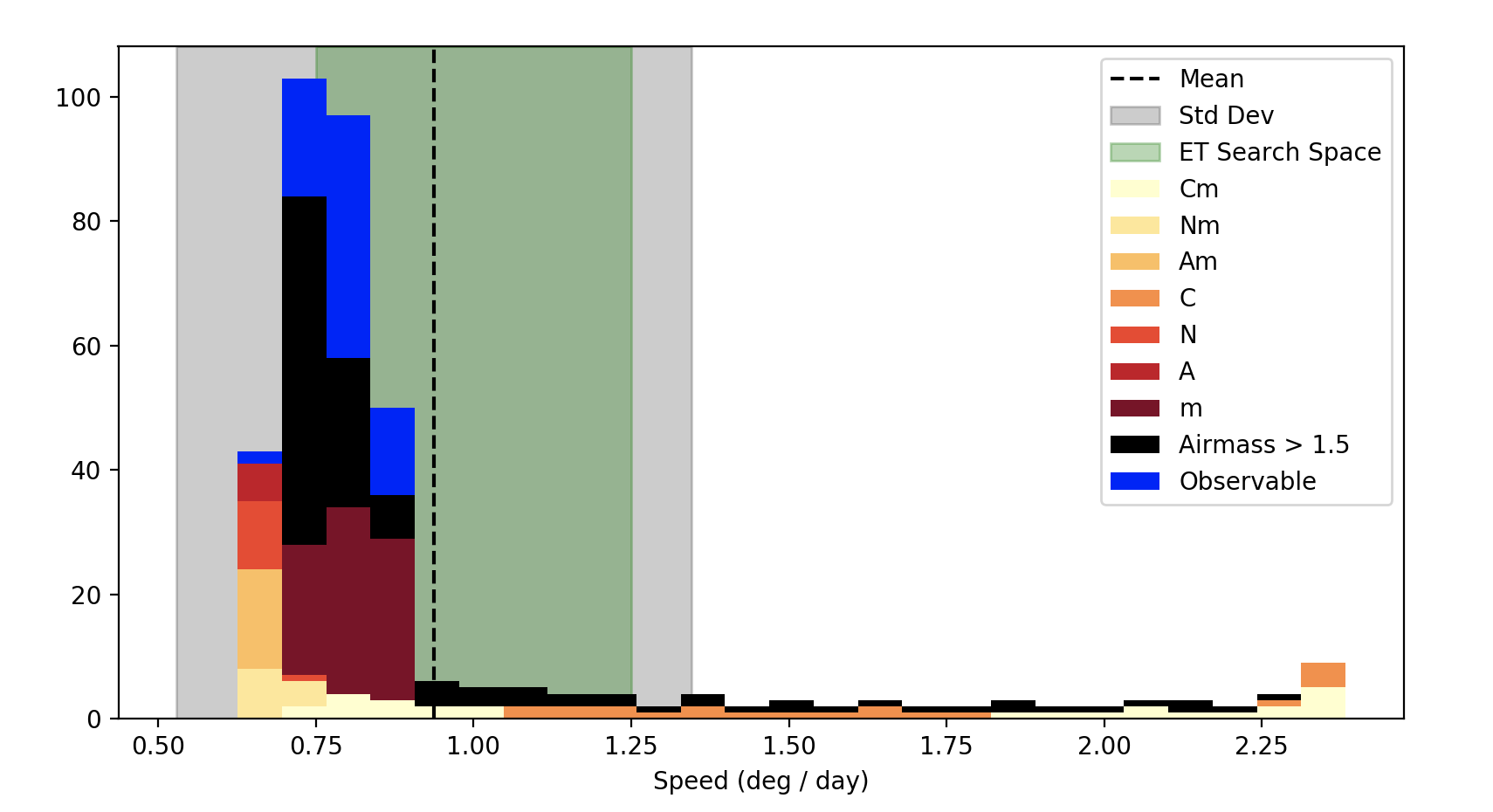}
  \caption{This figure shows the predicted rates of motion for \tk\ on each night in 2018 (taken from JPL HORIZONS: ssd.jpl.nasa.gov/?horizons). The dashed line shows the mean speed and the grey region depicts the standard deviation. The green region represents the speeds for which an object would be flagged as an ET by our pipeline. Nights on which we consider \tk\ to be observable are plotted in blue. The other colors represent nights on which observing \tk\ would be unfavorable according to the Solar Presence and Lunar Presence parameters from JPL HORIZONS. In particular, ``C" is civil twilight/dawn, ``N" is nautical twilight/dawn, ``A" is astronomical twilight/dawn, and ``m" is refracted upper-limb of Moon on or above apparent horizon. Although the standard deviation region extends beyond our assumptions for ET speed, the mean is well with our search space. In fact, most of the nights beyond our speed cuts are ones for which \tk\ is not observable.}
  \label{fig:tk7_speed}
\end{figure*}

\subsubsection{Flag ETs} \label{be_ets}

Due to the short arcs of our investigations, it was not possible to fit orbits to our detections, which would be required to definitively classify an object as an ET. However, due to their relatively consistent position with respect to the Earth and Sun, simulated ETs (see Sec.~\ref{fakes}) have a distinct rate of motion on the sky of 1 deg./day, except at small geocentric ranges (see Fig.~\ref{fig:et_speed}). Therefore, any previously undiscovered object in our catalog that has a mean speed between 0.75 and 1.25 deg./day is considered an ET candidate in our survey.

\par However, the only known ET, \tk, often has rates of motion that are not consistent with this assumption (see Fig.~\ref{fig:tk7_speed}). While technically its mean rate of motion is nearly 1 deg./day, it is clear that this object most often moves at $\sim$ 0.75 deg./day. Fortunately, this slower rate of motion often still falls within our speed assumption. Moreover, there are many evenings when \tk\ would not have actually been detectable due to high air mass, solar elongation, moon phase, etc. For a large portion of the evenings where \tk\ would have been detectable by our survey (if it had been directed towards L4), \tk\ would still have been flagged as an ET by our pipeline (if it had been detected). Regardless, \tk\ does not have an orbit that is representative of primordial ETs, and it should not be considered as a particularly strong or valid test for this pipeline. Thus, we consider our assumption of ET candidates having a speed between 0.75 to 1.25 deg./day to be reasonable since it is consistent with simulated ET populations and roughly valid for \tk\ (a notably atypical ET). 

\par However, an object that has a 1 deg./day rate of motion is \textit{not} guaranteed to be an ET. There are other populations of asteroids that can have similar rates of motion and could happen to overlap with the L5 point (see Sec.~\ref{aaobjects}). Again, additional observations of a candidate would be required to actually constrain its orbit and classify it as an ET.

\subsection{Verification by Eye/Catalog Creation}
\label{eye}

\par Despite the cuts on bad detections, transients, and links,  false positives were still found by the \textit{Bullseye} code. Therefore, linked objects were reviewed by hand in order to remove obvious false positives from our catalog. Such objects included detections of stationary objects with very variable astrometry between exposures, detections in the scattered light rings around very bright objects, detections along diffraction spikes, and detections along detector defects. Out of the 7676 candidate objects identified by the pipeline, 6059 were determined to be false positives. However, many of these anomalous objects were linked to large rings of scattered light around very bright or detector defects making them easy to identify and remove from our catalog.

\par While reviewing the candidate objects, it was also convenient to look for objects that were identified by SE and correspond to previously known Solar System objects but were not linked by the pipeline (i.e.~non-recovered objects). Many of these objects were observed to be in crowded field, confused with a bright star, or fell in a chip gap.  


\section{Detected Objects} \label{objects}

\subsection{Previously Known Objects}
\label{known_objects}

\begin{figure}
\includegraphics[width=0.5\textwidth,keepaspectratio=true]{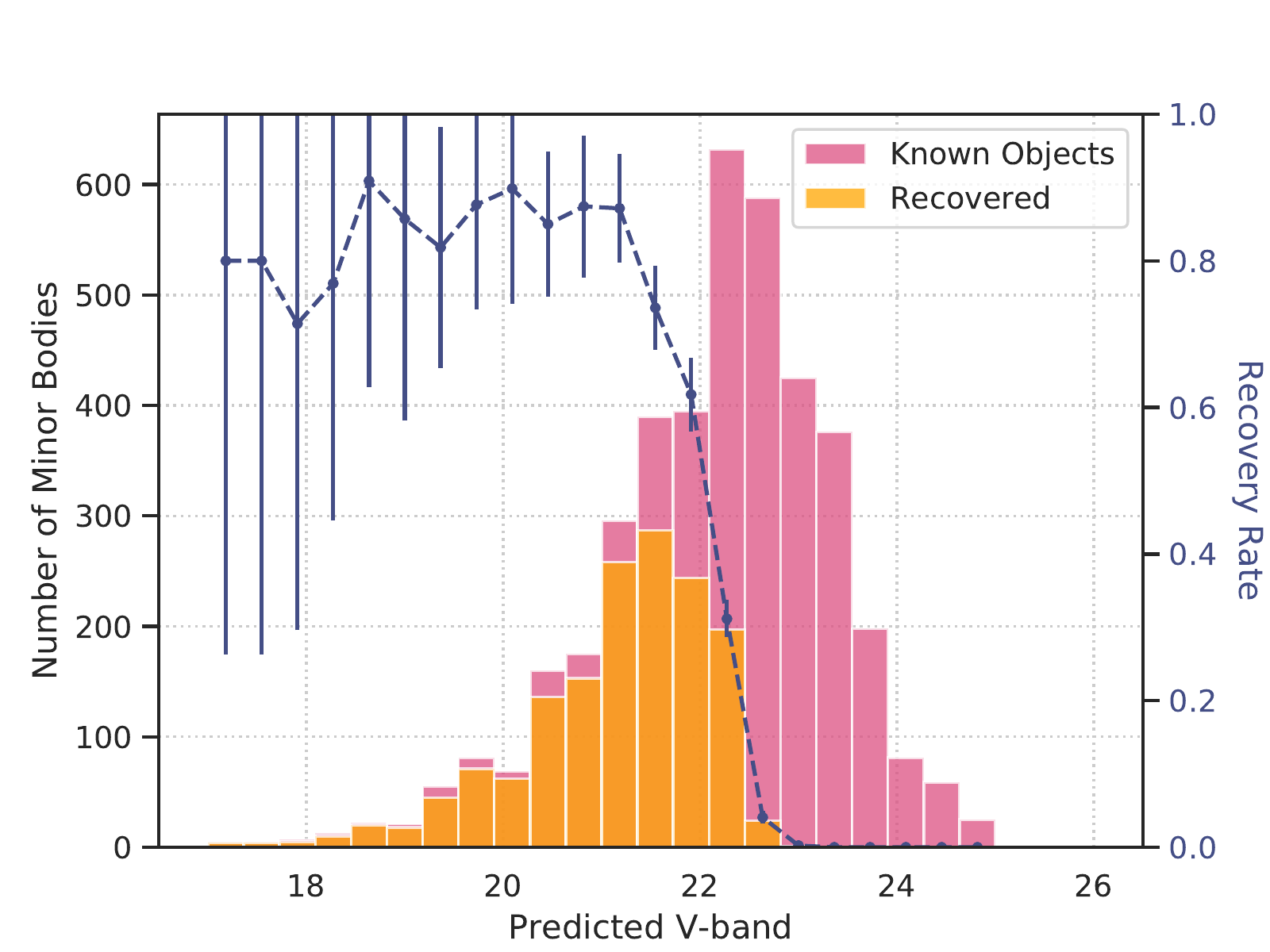}
\includegraphics[width=0.5\textwidth,keepaspectratio=true]{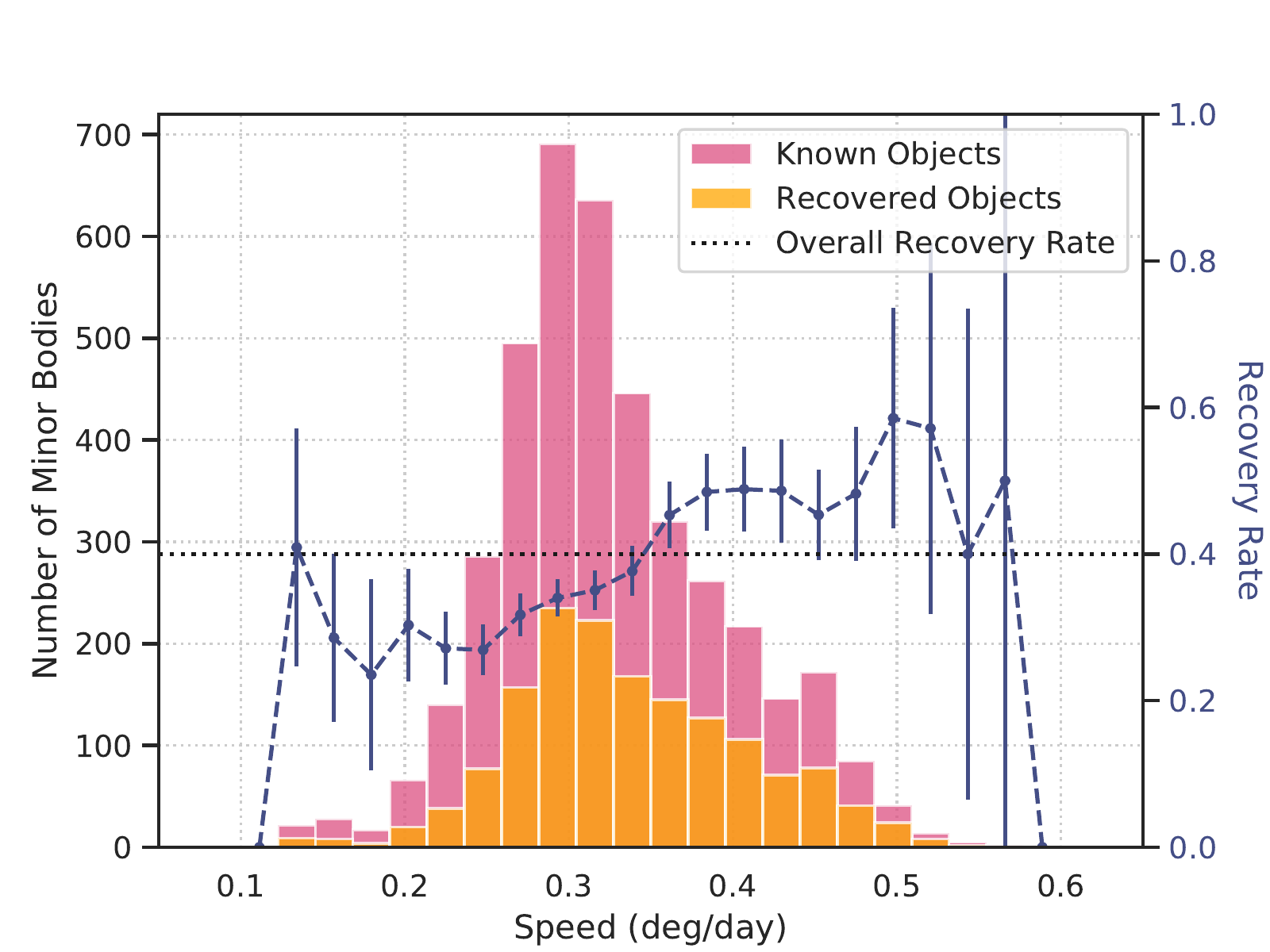}
  \caption[t]{\textit{Top}: Recovery rates for previously known minor bodies as a function of predicted V band magnitude. The known minor bodies that coincided with our survey area are shown in pink, and those which were detected and linked by our pipeline are shown in yellow. The recovery rate in each bin, with Poisson errors, are shown as a blue dotted line.
  \textit{Bottom}: This plot shows our recovery rates for previously known minor bodies as a function of predicted total sky motion. Colors are the same as \textit{Top}. The overall recovery rate of $\sim$40\% is shown as a black dotted line. The recovery rates across all of these speeds are generally consistent with this overall rate, but recovery rates are lower for slower speeds.}
  \label{fig:recovery_known}
\end{figure}

To determine which objects had been previously discovered, we compared our catalog to the SkyBoT database \citep{Berthier2006}. In total, there were 4126 minor bodies coincident with our survey area, and we recovered detections of 1546 known Solar System objects ($\sim$40\%). The majority of these objects were main belt asteroids. All of these detections have been submitted to the Minor Planet Center.\footnote{url{minorplanetcenter.net}}

\par Figure \ref{fig:recovery_known} shows our recovery of known objects as a function of their V-band magnitude as predicted by SkyBoT. As expected, our recovery rates drop off at fainter magnitudes. In particular, the recovery rates are $\gtrsim$~50\% for V $\lesssim$ 22, and they drop quickly to 0\% beyond this point. This implies that overall limiting magnitude for our survey is roughly V = 22. However, reducing this limit to a single number obfuscates the different detection thresholds in each image due to differences in chips, weather conditions, etc. This limiting magnitude also explains why the overall recovery rate for our survey was only 40\% as a majority of the previously known objects were not detectable in our survey. 

\par Fig. \ref{fig:recovery_known} also shows our recovery rates as a function of their rate of motion as predicted by SkyBoT. The recovery rates in each speed bin are roughly consistent with the overall recovery rate. Therefore, we conclude we are not especially more/less sensitive to detecting objects moving at a particular speed. However, these recovery rates are not entirely independent of speed; slow moving rates are somewhat lower than for fast moving objects. This discrepancy is likely due to the fact that slower moving objects are more likely to be flagged as stationary. Since ETs move at a speed of $\sim$ 1 deg. / day, they were very unlikely to be flagged as stationary in our dataset and, as is shown in Fig. \ref{fig:recovery_known} we may even be more efficient such objects.
 
\par Moreover, due to the nature of our survey, it is likely that we observed these objects at much higher phase angles than their discovery. Therefore, we expect our recovery rate and limiting absolute magnitude of known asteroids to be lower than an identical survey of objects at lower phase angles.

\begin{table}
\caption{New Objects. The individual observations for all of these objects have been submitted to the \textit{Minor Planet Center}.  Columns: (1) Our designation for this object. The first number is the field and the second number is the chip the object was found in. The third number is a unique id; (2) Inferred V-band magnitude for the object. See Sec.~\ref{fakes} for a description of how the V-band magnitudes were calculated; (3) Rate of sky motion measured for the object (deg./day). The rate of motion is averaged over all of the detections; (4) Position angle measured for the object (deg.). The P.A.~is averaged over all of the detections; (5) Number of detections of the object.}
\begin{center}
\begin{tabular}{lrrrr}
\hline
Designation & m$_V$ & Rate of Motion & P.A. & Num.\\
\hline
  1\_55\_5 &  23.1 &   0.29 &  129.3 &    6 \\
  2\_37\_0 &  22.5 &   0.37 &  115.3 &    6 \\
  2\_51\_2 &  22.6 &   0.40 &   99.9 &    7 \\
  2\_54\_9 &  22.8 &   0.37 &  113.3 &    5 \\
  3\_21\_2 &  23.0 &   0.49 &  110.9 &    6 \\
  3\_34\_3 &  23.5 &   1.76 &  264.3 &    5 \\
 4\_15\_13 &  22.6 &   0.15 &  122.3 &    7 \\
  4\_36\_0 &  22.8 &   0.32 &  113.1 &    8 \\
  4\_39\_6 &  23.4 &   0.27 &   99.7 &    5 \\
  4\_45\_0 &  22.9 &   0.49 &  107.9 &    7 \\
  4\_56\_5 &  23.3 &   0.14 &  129.1 &    5 \\
  5\_16\_2 &  22.8 &   0.46 &  107.5 &    7 \\
  5\_32\_1 &  23.2 &   0.32 &  102.5 &    6 \\
  6\_24\_5 &  23.3 &   0.36 &  115.7 &    6 \\
  6\_34\_0 &  22.9 &   0.32 &  121.5 &    8 \\
  6\_58\_2 &  22.4 &   0.36 &  153.8 &    7 \\
   7\_3\_6 &  23.3 &   0.42 &  107.9 &    5 \\
   7\_4\_3 &  23.1 &   0.30 &   89.1 &    6 \\
  7\_15\_1 &  23.0 &   0.32 &  130.9 &    6 \\
  7\_18\_3 &  22.8 &   0.16 &   90.2 &    8 \\
  7\_19\_2 &  22.9 &   0.36 &  105.3 &    7 \\
  7\_19\_3 &  22.8 &   0.49 &  106.3 &    6 \\
  7\_20\_3 &  23.2 &   0.38 &  113.9 &    6 \\
  7\_27\_2 &  22.6 &   0.33 &  109.5 &    8 \\
  7\_33\_5 &  23.2 &   0.39 &  114.3 &    5 \\
  7\_37\_0 &  22.1 &   0.26 &  104.0 &    8 \\
  7\_50\_1 &  22.6 &   0.33 &  129.3 &    7 \\

\end{tabular}
\end{center}
\label{table:new_objects}
\end{table}

\subsection{New Objects}
\label{new_objects}

\par Table \ref{table:new_objects} lists all of the Solar System objects identified by our pipeline that did not match to positions of known asteroids (a total of 27 new objects). Most importantly, none of these objects have a rate of motion of $\sim$ 1 deg./day. Rather, they are consistent with the rates we observed for main belt asteroids. Therefore, by our definition, we did not find any ET candidates in our survey. However, due to the short arcs of our dataset, we cannot determine the orbits of these new objects or their classification. These detections have been submitted to the Minor Planet Center (MPC) so that they may be linked to an orbit in the future. 

\begin{table*}
\caption{Previously Known Objects Flagged as ETs. The individual observations for all of these objects have been submitted to the \textit{Minor Planet Center}. Columns: (1) Object Designation; (2) Orbit type taken from \textit{SkyBoT} \citep{Berthier2006}; (3) Predicted V-band magnitude (taken from \textit{SkyBoT}); (4) Rate of sky motion measured for the object (deg./day). The rate of motion is averaged over all of the detections; (5) Position angle measured for the object (deg.). The P.A.~is averaged over all of the detections; (6) Number of detections of the object. 
}
\begin{center}
\begin{tabular}{llrrrr}
\hline
Designation & Orbit Type & m$_V$ & Rate of Motion & P.A. & Num. Detections \\
\hline
2003 HF2 &  NEA:Apollo &  21.9 &   1.08 &  110.5 &    6 \\
2006 JF42 &    NEA:Aten &  21.0 &   1.17 &  124.1 &    7 \\

\end{tabular}
\end{center}
\label{table:flagged_objects}
\end{table*}

\subsection{Aten \& Apollo Asteroids}
\label{aaobjects}

While our pipeline did not flag any new objects as ET candidates, there were a few previously known objects that were flagged because they happened to have a rate of motion of $\sim$ 1 deg/day (see Table \ref{table:flagged_objects}). One of these objects was an Apollo asteroid while the other was an Aten. We draw attention to these objects specifically because they provided a serendipitous test of our pipeline while also highlighting the limitations of our method. 

\par
\begin{figure*}
\includegraphics[width=\textwidth,height=\textheight,keepaspectratio=true]{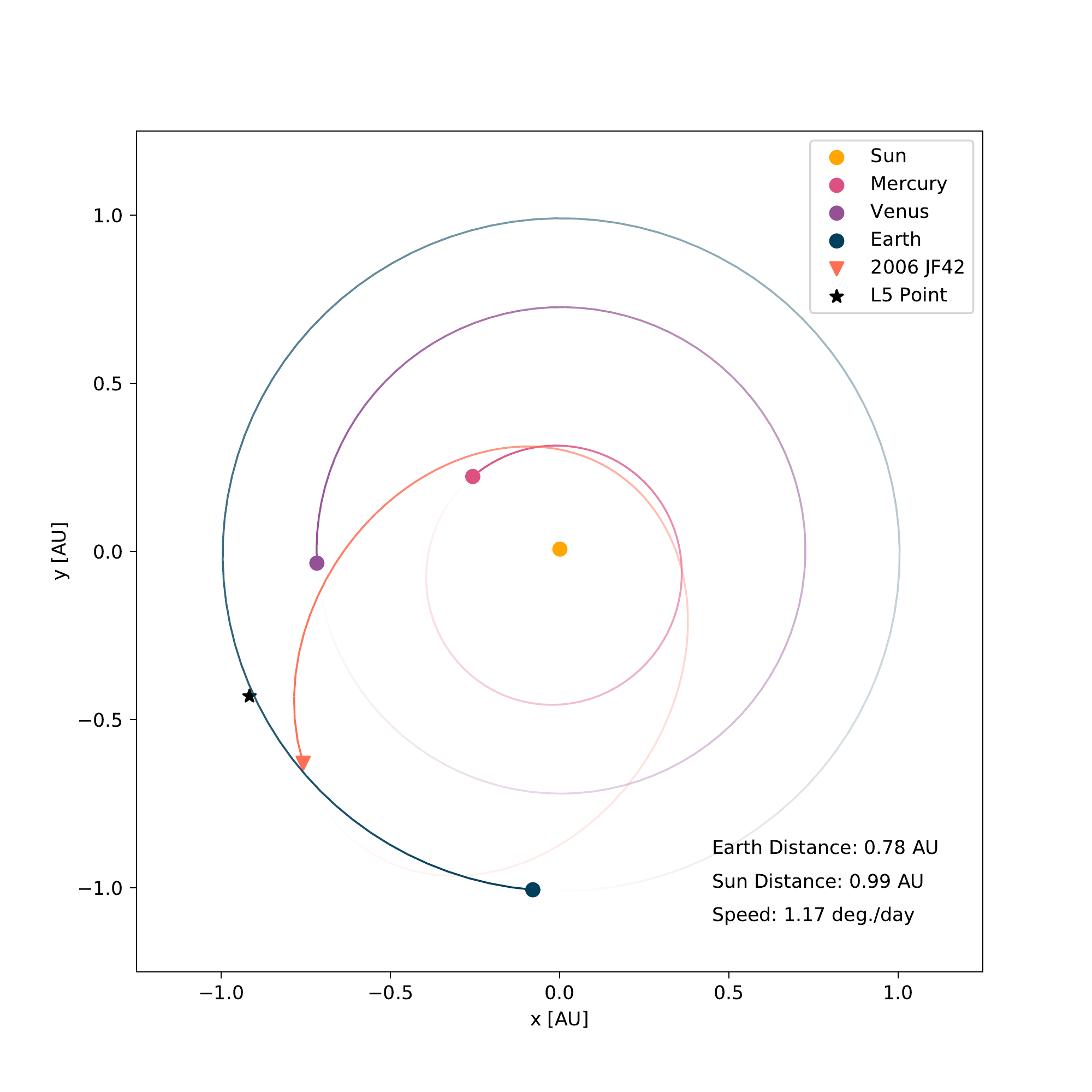}
  \caption[t]{This figure shows the orbit of the Aten asteroid detected in our survey, $2006~JF_{42}$. At the time of our observations, this asteroid was near the L5 point. With a distance to the Earth of 0.78 AU and a distance to the Sun of 0.99 AU, this object had a rate of motion of 1.17 deg./day, causing it to be misidentified as an ET candidate by our pipeline. This figure was produced using REBOUND \citep{Rein2011}.
  }
  \label{fig:apollo}
\end{figure*}

 First, Apollo and Aten asteroids both have Earth crossing orbits, with q \textgreater 1 AU and q \textless 1 AU respectively. If these asteroids were to cross the Earth's orbit at a point near the L4 or L5 points with a rate of motion of $\sim$ 1 deg./day they would be indistinguishable from ET candidates in our survey. 
Fig.~\ref{fig:apollo} shows the orbit of the flagged Aten asteroid, which indeed crosses the Earth's orbit near the L5 point and has a rate of motion of 1.17 deg./day. The orbit for the flagged Apollo asteroid is similar. The fact that these distinct populations would be indistinguishable in our observations highlights the fact that our method can \textit{not} definitively classify transient objects as ETs. 

\par However, it is still significant that these objects were flagged. Since these asteroids behave exactly as we expect ETs would in our dataset and they were properly detected and flagged, this implies that our pipeline was capable of finding ETs as well. Certainly, these objects are not a robust test of the capability of the pipeline, but they did allow for a test on real, known objects. The only known ET that we could have otherwise tested on (\tk) is at the L4 point, which is not within our survey area. 

\section{Upper Limits on ET Population} \label{limits}

\par To calculate an upper limit on the ET population, we generated a population of simulated ETs, injected them into our data set, ran these images through the pipeline as described in Section \ref{pipeline}, and then measured our recovery rate for these objects. We then used this recovery rate to place an upper limit on the ET population. 

\subsection{Synthetic ETs}
\label{fakes}

To generate synthetic ET populations, we adopt the restricted three body model of Sun, Earth and a test particle. We generated two different synthetic ET sets: a ``Strict" ET set and a ``Jupiter Trojan" ET set. Both sets were used to predict the rate of motion for ETs (see Sec.~\ref{be_ets}), but only objects from the Strict set were injected into our images.

\subsubsection{Strict ET Set}
We generated a sample of $N=10^6$ Earth-like test particle orbits, with heliocentric osculating orbital elements as follows: semi-major axes uniform random in the range $0.99-1.01$ AU, eccentricities uniform random in $0-0.1$, ecliptic inclinations uniform random in $0-0.1$ radians; the angular elements (longitude of ascending node, argument of perihelion and true anomaly) uniform random values in the range $0-2\pi$.  We computed the position and velocity vectors, ${\bf r}$ and ${\bf v}$, in the barycentric frame, and the Jacobi integral for these particles,
\begin{equation}
\label{renu_1}
	C = 2( {1-\mu)\over |{\bf r}-{\bf r}_\odot|} + {\mu\over |{\bf r}-{\bf r}_\oplus|}) +2\hat{\bf n}\cdot ({\bf r}\times{\bf v})  -v^2
\end{equation}
where $\mu$ is Earth's mass as a fraction of the total mass (Earth and Sun) and $\hat{\bf n}$ is the unit vector normal to the ecliptic. 
We then selected the subset of those particles having Jacobi integral in the range $3-\mu$ to $3+\mu$, in standard units for the restricted three body problem; this range of Jacobi integral covers the L4 and L5 libration zone. The synthetic L4 and L5 ETs were then selected as the subset having instantaneous positions within 30 degrees of heliocentric ecliptic longitude of the L4 point and L5 point, respectively. This produced a sample of 3797 L4 and 3726 L5 synthetic ETs.

\subsubsection{Jupiter Trojan ET Set}

We started with the full dataset of minor planet orbital data from the Minor Planet Center (on Nov 27, 2018). We selected the observationally complete set of Jupiter Trojans from this set, which we defined as those objects of absolute magnitude H \textless\ 13.5 and semi-major axis in the range $5.0-5.4$ AU (2844 object in total). To generate ET orbits, we made three changes to the orbital data set of Jupiter Trojans. First, we rescaled the semi-major axes of the particles to Earth's,
\begin{equation}
\label{renu_1}
	a_{ET} = a_{\oplus} + (a-a_{Jupiter})(R_{H,\oplus}/R_{H,Jupiter})
\end{equation}
where $R_{H,\oplus}, R_{H,Jupiter}$ are the Hill radii of Earth and Jupiter, respectively. Second, we computed the orbital plane orientation of each Jupiter Trojan relative to Jupiter's osculating orbital plane and re-assigned it to be relative to Earth's osculating orbital plane. Third, we assigned the mean anomaly of ETs as follows:
\begin{equation}
\label{renu_3}
	M_{ET} = M + \lambda_{\oplus}-\lambda_{Jupiter}
\end{equation}
where $\lambda_{\oplus}, \lambda_{Jupiter}$ are the mean longitudes of Earth and Jupiter at the epoch of the Minor Planet Center's data, and $M$ is the mean anomaly of a Jupiter Trojan. These changes produce a population of synthetic ET orbits that has the same dispersion in semi-major axis, eccentricity and inclination as the Jupiter Trojans, but relative to Earth's orbit and scaled to the size of Earth's Trojan regions.

\begin{figure}
\includegraphics[width=0.5\textwidth,keepaspectratio=true]{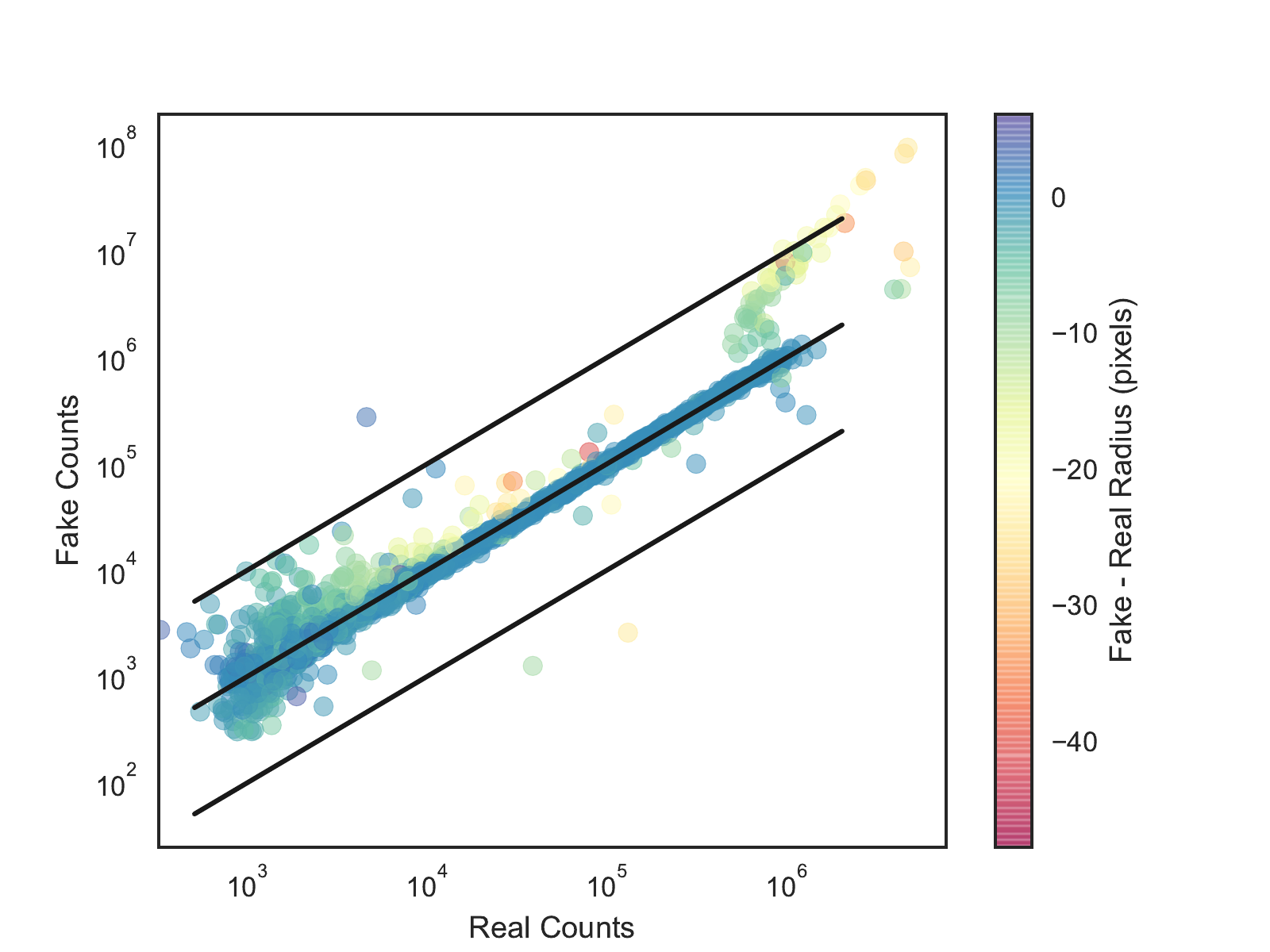}
\includegraphics[width=0.5\textwidth,keepaspectratio=true]{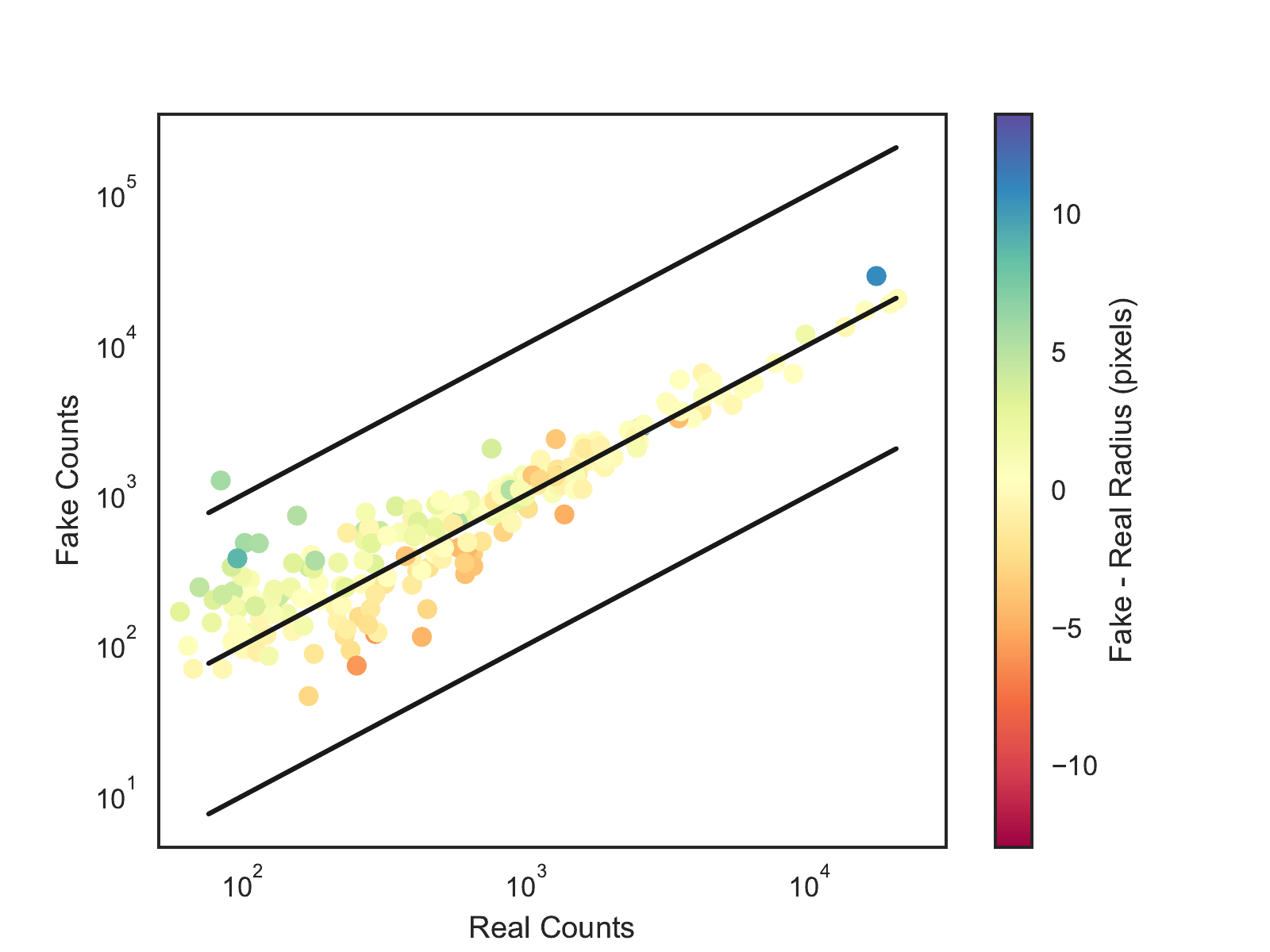}
  \caption[t]{\textit{Top}: This figure compares the number of counts measured by SE for Solar-type stars in Field 0 to the number of counts for a fake source of the same V-band magnitude. In general, the points lie well along the one-to-one line and the radii of the fake and real sources are nearly the same. Not surprisingly, there is more dispersion from this line at the faint end, but the points still have the same number of counts within an order of magnitude. There is also a clear overabundance of fake counts for the brightest objects, which is likely due to the saturation limit. \textit{Bottom}: Same as \textit{Top} but for known Solar System objects in Field 0. These points fall well along the one-to-one line and the radii of the real sources as compared to the fake ones are still nearly the same. Based on this plot, we conclude that our fake sources are consistent with our observations of known Solar System objects in our survey.  
  }
  \label{fig:counts_mags}
\end{figure}
\begin{figure}
\includegraphics[width=0.5\textwidth]{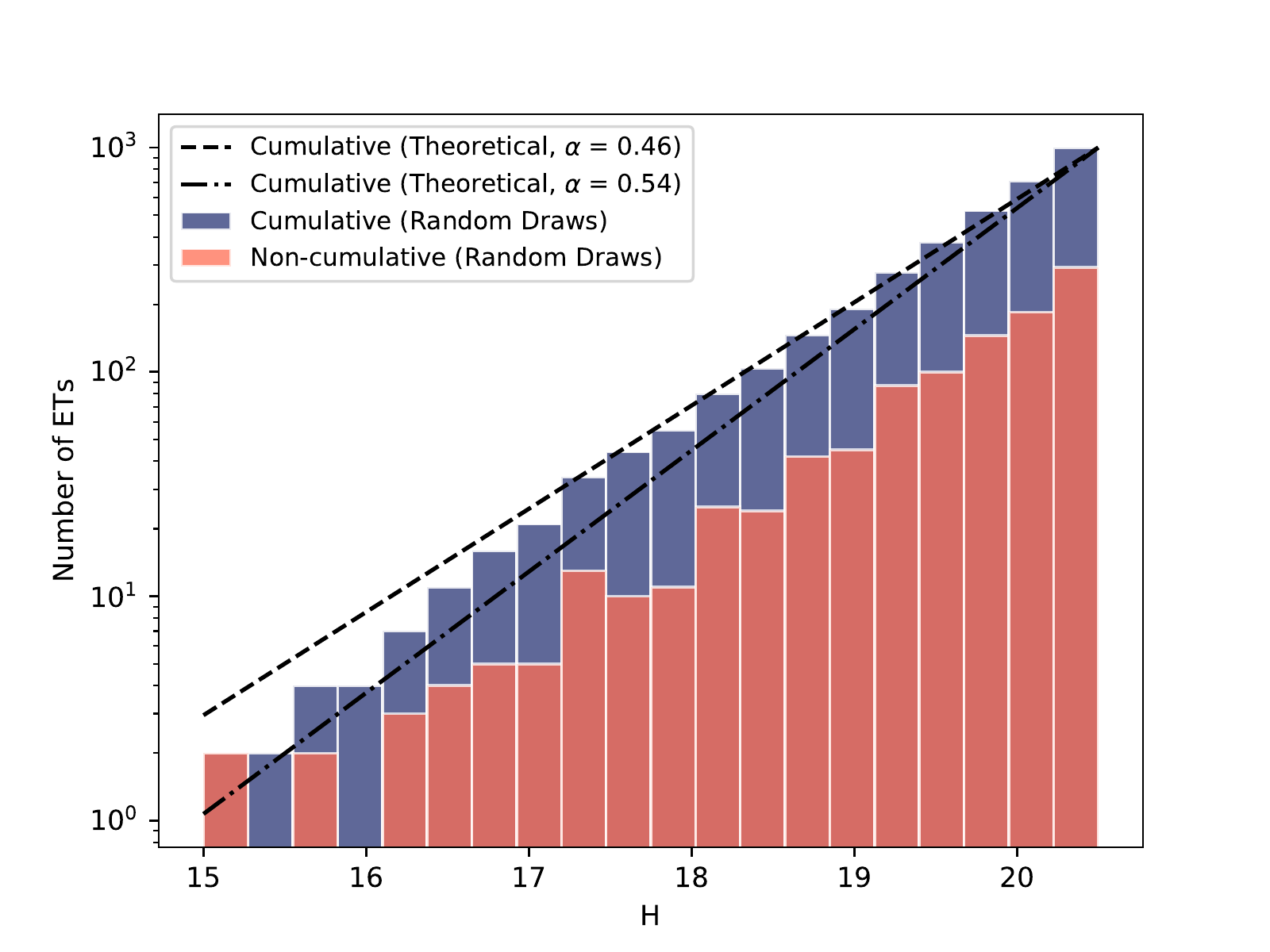}
  \caption[t]{This figure compares our theoretical H distribution of ETs (Eq.~\ref{mag_dist}) to the distribution generated by random draws from Eq.~\ref{mag_dist} normalized by the total number of fake ETs (in the case of this figure, $10^3$). The black dashed and dot-dashed lines represent the theoretical H distribution with $\alpha = 0.46$ and 0.54, respectively. The blue and red histograms represent the cumulative and non-cumulative H distributions, respectively, which were generated for the simulated ETs (see Sec.~\ref{inject} for description). Due to the random nature of this method, these distributions are similar to those used for the sources that were actually injected into our images to asses our recovery rate, but they are not exactly the same. Most importantly, the simulated cumulative distribution lies well between the two theoretical distributions. We therefore conclude that the injected ETs have an H (and thus size) distribution consistent with what we would expect theoretically.}
  \label{fig:h_dist}
\end{figure}
\subsection{Injection of Synthetic ETs}
\label{inject}

To estimate our recovery rate for ETs, we drew from the Strict synthetic ET set and injected corresponding sources into our images.\footnotemark[1] To most accurately simulate the ET population in our images, the synthetic sources must be at realistic positions, follow a reasonable absolute magnitude (H) and physical size distribution, and have the correct amount of injected counts for the corresponding magnitude. 

\par The synthetic sources were injected as Gaussian point sources into the image array count by count. Note, assuming an albedo of 0.15, we expect ETs to be less than about 1 km in size; such objects would not be resolved in our dataset, so we can safely model them as point sources. Each synthetic source is generated to have the same FWHM as derived by the DECam Community pipeline for each image. In particular, the width $\sigma$ of a Gaussian distribution can be derived from the FWHM via the relation 
\begin{equation}
\label{fwhm_sigma}
    FWHM = 2 \sqrt{2 \ln 2} \sigma
\end{equation}
Each count was added at a random x, y shift from the center of the object generated from this distribution. This method worked well for many sources, but not all; very bright sources often had a much larger angular size than the inferred FWHM (see Fig.~\ref{fig:counts_mags}).

\par To calculate the sky positions of our synthetic ETs, we used the generated heliocentric osculating orbital elements (see Sec.~\ref{fakes}) and \textit{PyEphem}\footnote{url{https://rhodesmill.org/pyephem/}} to calculate the predicted RA, Dec at the time of each of our observations. Our observing location at Cerro Tololo Inter-American Observatory (longitude: -70.80\textdegree, latitude: -30.17\textdegree, elevation: 2207 m) was taken into account for these calculations. 

\par For the H distribution, we use one similar to that of NEOs/asteroids:
\begin{equation}
\label{mag_dist}
    N(<H) = A * 10 ^ {\alpha H}
\end{equation}
where H is the absolute magnitude, A is a normalization factor, and $\alpha$ is a slope parameter. We draw uniformly from a range of typical $\alpha$ = 0.46 - 0.54 \citep{JEDICKE1998245, Bottke2190, HARRIS2015302, JEONGAHN2015}. We set A such that N(\textless  H\textsubscript{max}) is equal to the total number of synthetic ETs. Of course, this means that we need to pick an H\textsubscript{max}; we defined our distribution to have H\textsubscript{min} = 15 and H\textsubscript{max} = 20.5. Our choice of H\textsubscript{min} is somewhat arbitrary as very bright objects will be very rare in this distribution. Based on our recovery of previously known Solar System objects (see Sec.~\ref{known_objects}), we estimated our limiting magnitude to be $\sim 21.5$ in V-band, but we still recovered known objects at V $\sim 22.5$ (see Sec. \ref{known_objects}). This magnitude corresponds to H $\sim 20.4$. Therefore, we inject synthetic ETs up to H\textsubscript{max} = 20.5.

\par To generate a magnitude for each specific object, we cannot use Eq.~\ref{mag_dist} directly as it is a cumulative distribution. Instead, we normalize Eq.~\ref{mag_dist} by A, such that N(\textless  H\textsubscript{max}) = 1 and then use this distribution as a probability distribution. 
Fig.~\ref{fig:h_dist} demonstrates that this method leads to magnitudes that are consistent with the theoretical distribution (i.e.~the cumulative distribution of our randomly drawn magnitudes matches the theoretical distribution well). Therefore, we conclude that this method is sufficient to generate synthetic ETs with an accurate distribution. 

\par Since each object's apparent magnitude is highly dependent on its location in its orbit, we use the functionality in \textit{PyEphem} to calculate an apparent magnitude for the synthetic ETs at the time of each observation. To convert these magnitudes into counts, we assume a linear relationship between counts and the flux of the source. This function will be slightly different between each chip as they are not all identical. Therefore, chip by chip, we identify SDSS sources
in each image using \texttt{astroquery}\footnote{url{doi.org/10.5281/zenodo.1160627}} and then fit a line to their counts as measured by SE (see Sec.~\ref{se}) vs. the apparent V band magnitude . We calculated the V band magnitude from the SDSS g and r magnitudes using 
\begin{equation}
\label{g_r_to_v}
   V = g - 0.59 * (g -r) - 0.01
\end{equation}
\citep{Jester_2005}.
We used V-band magnitudes in order to compare to the predicted V-band magnitudes from SkyBoT \citep{Berthier2006} for the known Solar System objects in our image. 

\par One may expect that our fitted lines should have slopes of -0.4. However, we found that restricting our fit to have a slope of this value led to synthetic sources with counts that were systematically higher than the real sources.  
Since the light observed from asteroids in optical wavelengths is primarily reflected sunlight, we compared the fitted line to only SDSS sources with solar-like colors: g-r = $+0.44\pm0.02$. 
We found that these fits consistently overpredicted the number of counts for faint sources. We suspect that this discrepancy is due to a breakdown in the assumption that VR and V magnitudes should be comparable. In particular, the VR filter is essentially a very broad r filter ($\lambda \sim 475 - 750$~nm) that overlaps with the g band. Since solar-type objects are more red (g-r = +0.44), this implies that the the flux overlapping with the g-band would be lower than expected assuming  the object had a flat spectrum over VR. Therefore, the overlap with the g-band would be overpredicted, leading to consistently higher flux counts overall for solar-type objects. Therefore, we instead fit a line to only the solar-type SDSS objects in our images without imposing a slope constraint. 

\par Fig.~\ref{fig:counts_mags} shows the final comparison between the real counts for SDSS sources and Solar System object sources versus the measured counts for their synthetic counterparts. In both cases, most of the sources follow the one-to-one line, and nearly all of the synthetic sources have a number of counts within an order of magnitude of the real counts. There is more deviation at the faint end, but this is expected since this regime is more dominated by noise. There is a noticeable systematic surplus of counts for synthetic SDSS objects at the bright end (Fig.~\ref{fig:counts_mags}). This discrepancy is likely due to the saturation limit; while real sources cannot have a number of counts exceeding this limit, there is no such restriction for the synthetic objects, allowing them to have increasingly more counts as compared to the real sources. Since we do not see this problem at the faint end or for the Solar System objects, we do not attempt to correct for it. Also, in most cases, the angular size of the synthetic sources is nearly the same as the real source. Since the synthetic objects have similar sizes and counts compared to the real sources, we conclude that our method is able to inject realistic synthetic objects into our data set. 

\begin{figure*}
\includegraphics[width=\textwidth,keepaspectratio=true]{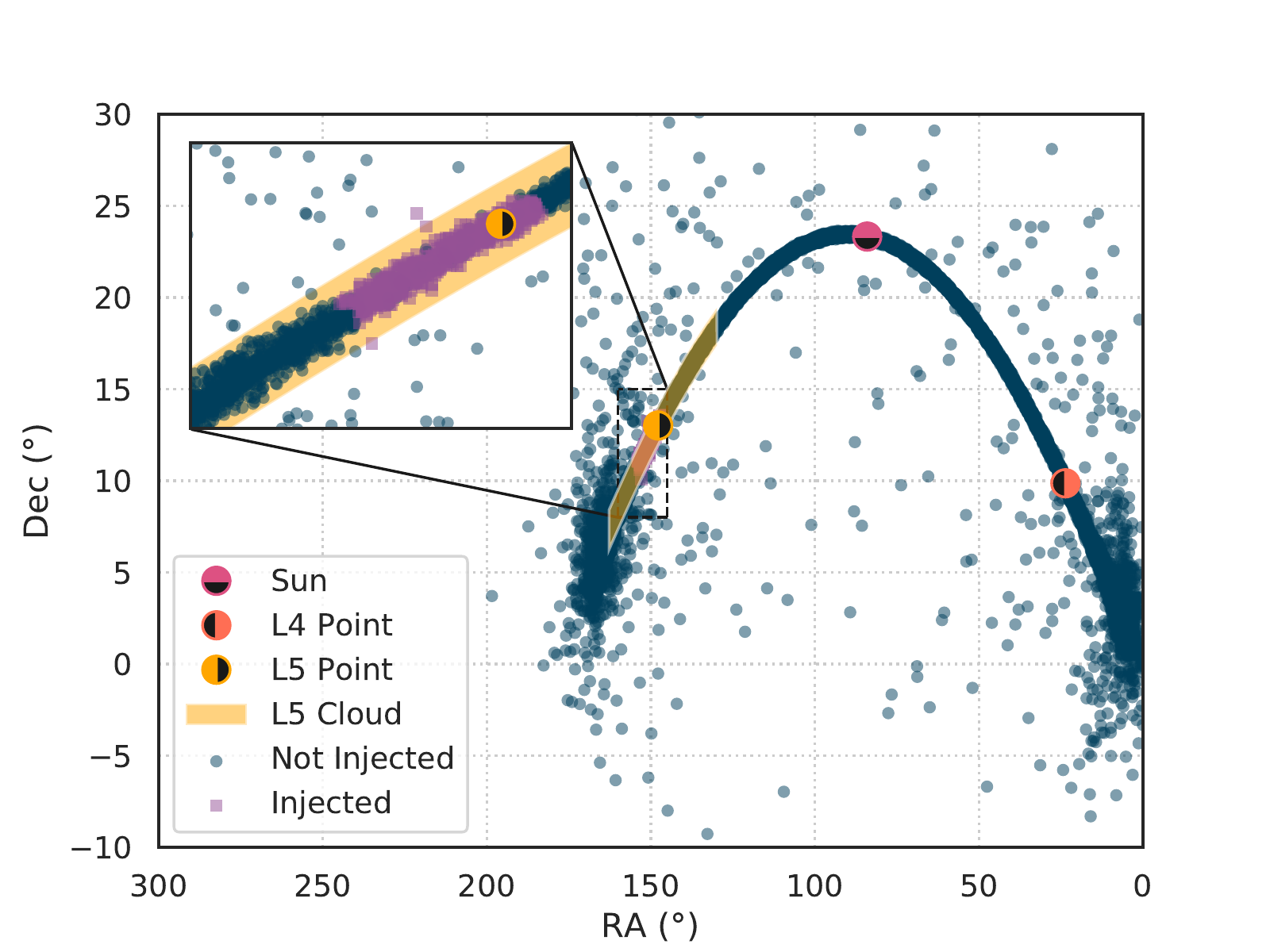}
  \caption[t]{This figure demonstrates coverage by our survey of the synthetic ET population. The Lagrange points and the Sun are plotted as half filled circles for reference. The yellow shaded area represents the L5 cloud (the area with ecliptic longitude within 30 degrees of L5). Synthetic ETs (see Sec.~\ref{fakes}) that were injected into our images are plotted as purple squares, while those that were not are blue circles. These injected points essentially depict the area covered by our survey. These objects make up 24\% of the ETs in the L5 cloud.}
  \label{fig:inject_pos}
\end{figure*}
\begin{figure}
\includegraphics[width=0.5\textwidth,keepaspectratio=true]{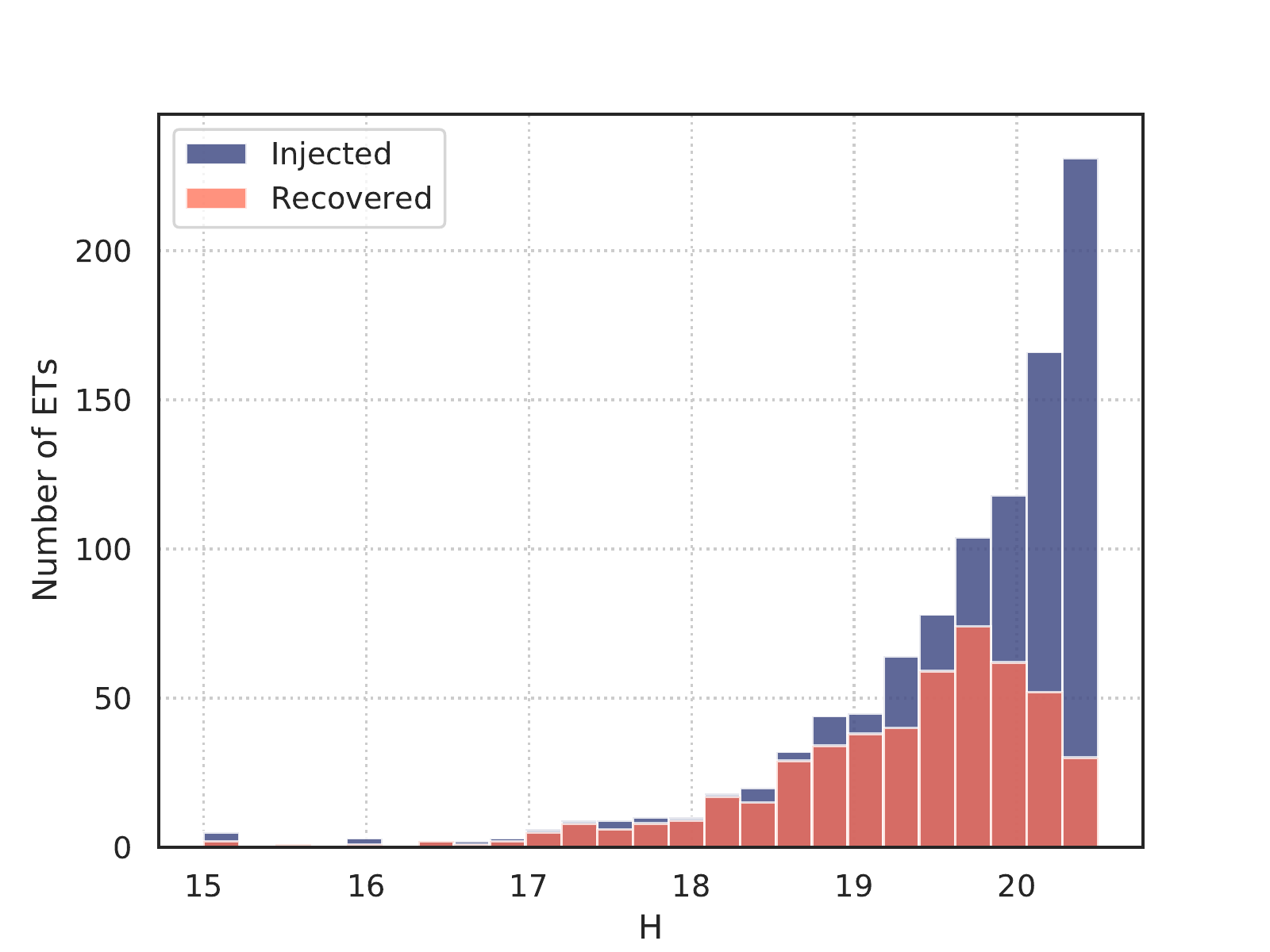}
  \caption[t]{This figure compares the injected ETs (blue) from the set of Synthetic ETs to those that were detected, linked, and flagged as ET candidates by our pipeline (red); in other words, this plot shows our recovery rates as a function of H. We have nearly 100\% recovery rates for most of the bins at the bright end, and the rates begin to fall off at $\sim$H=19.7. This value is essentially our limiting magnitude in H (though this does not directly relate to our limiting magnitude in V-band). Overall, we recovered $\sim$50\% of the injected ETs.}
  \label{fig:recovery}
\end{figure}
\begin{figure}
\includegraphics[width=0.5\textwidth, keepaspectratio=true]{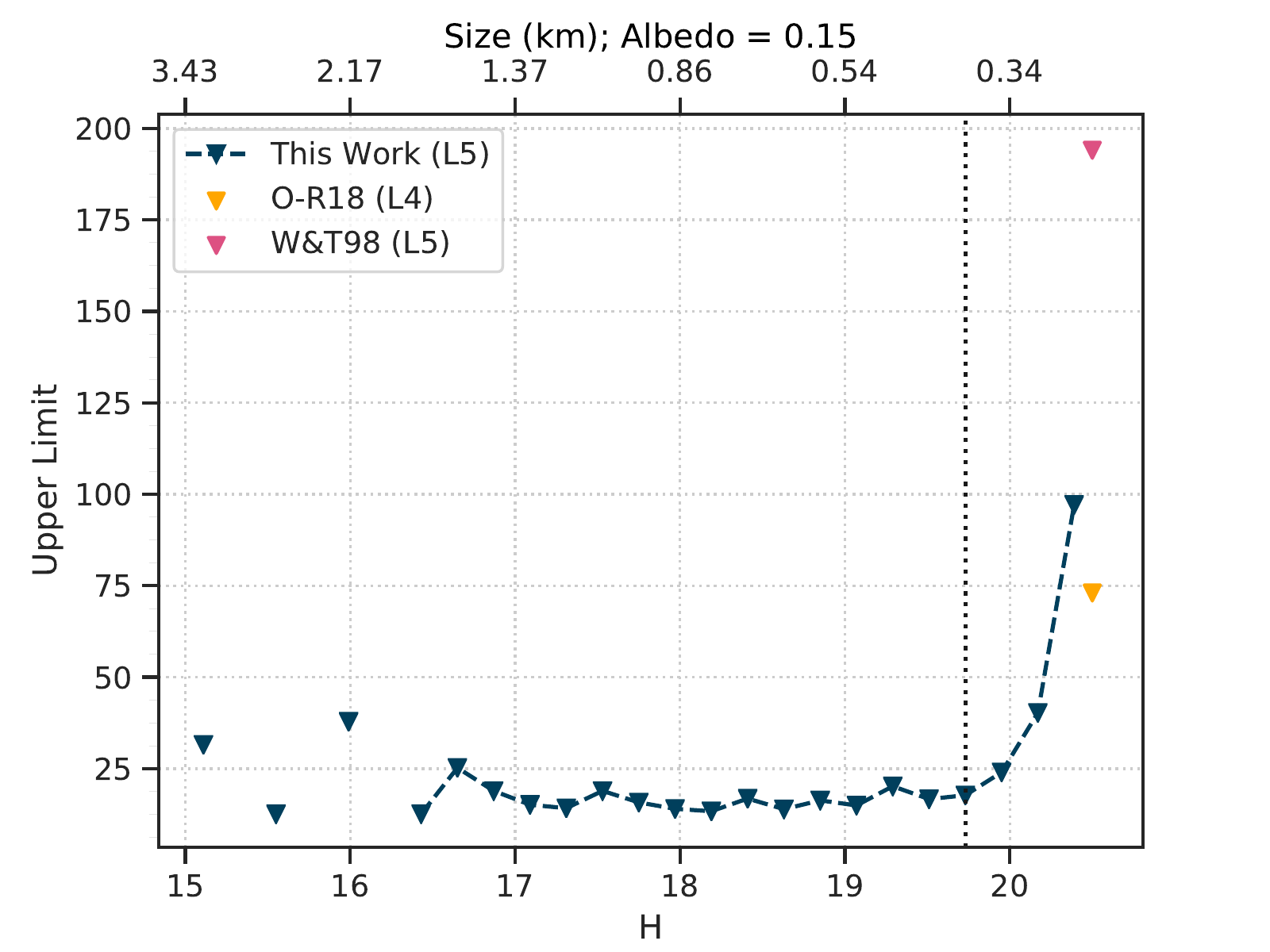}
  \caption[t]{Upper limits on the ET population calculated from our survey area coverage (Fig.~\ref{fig:inject_pos}) and recovery rate (Fig.~\ref{fig:recovery}). The yellow triangle depicts the upper limit calculated for L4 ETs by \cite{Cambioni2018}, while the pink triangle is the upper limit extrapolated from the \cite{Whiteley1998} results by \cite{Cambioni2018}. The upper limits from this work have two distinct regimes: flat at the bright end and steeply increasing at the faint end. We get a sharp increase for the faintest magnitudes due to our limiting magnitude and relatively poor recovery rates for faint ETs. For the faintest bin, H=20.4, we calculate an upper limit of 97 ETs. This limit is clearly more stringent than that from the \cite{Whiteley1998} results (194) and somewhat higher than the \cite{Cambioni2018} limit (73) for L4. Our recovery rates are flat at the bright end because all of these bins had nearly 100\% recovery rates leading to a constant upper limit ($\sim$15) for those magnitudes. The dotted line represents the faintest magnitude (H=19.7) still within the flat regime. At this magnitude, we calculate an upper limit of 18.}
  \label{fig:upper_lim}
\end{figure}
\begin{figure}
\includegraphics[width=0.5\textwidth,keepaspectratio=true]{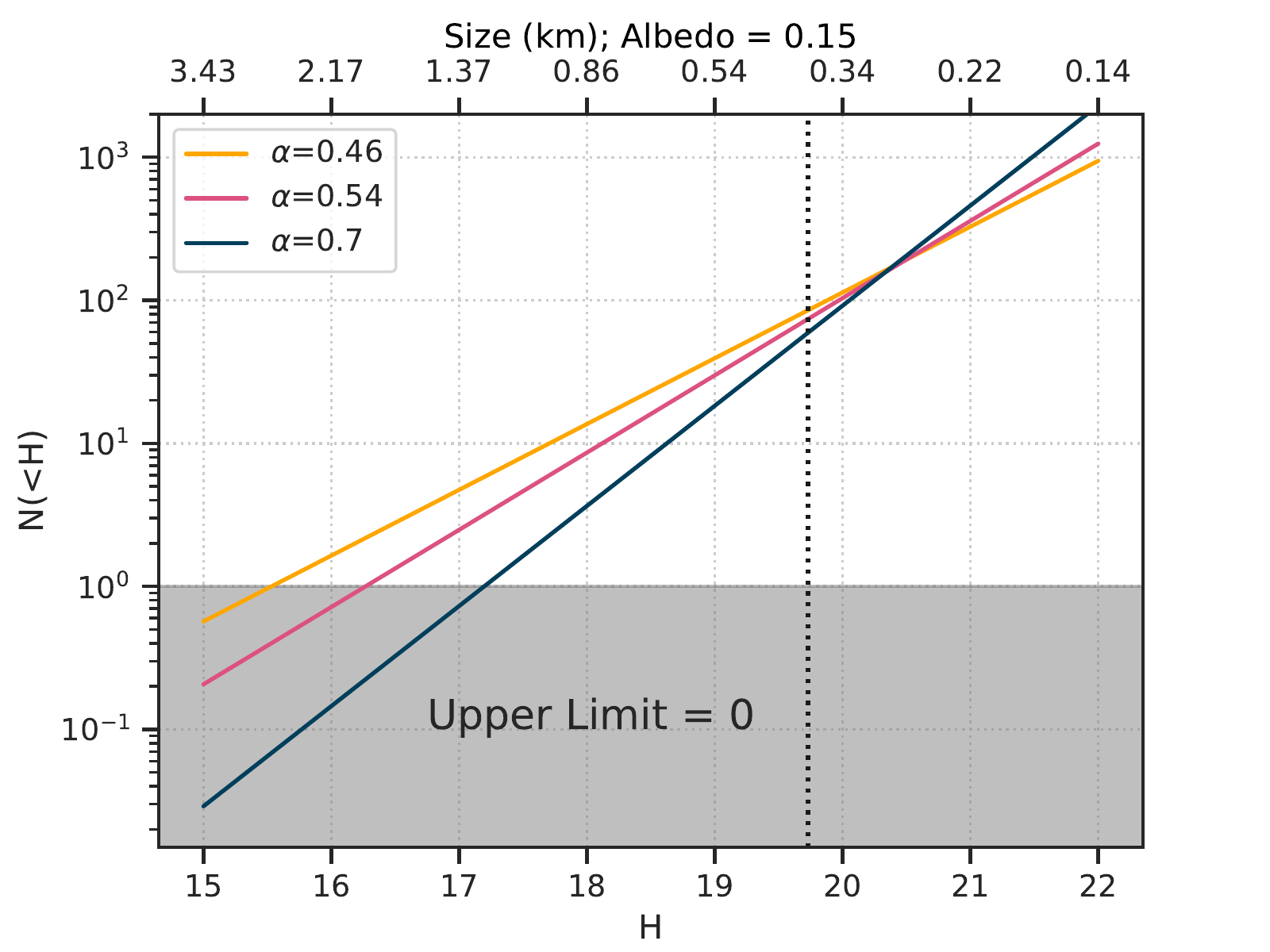}
  \caption[t]{This figure depicts the stricter upper limits calculated for bright ETs assuming the ET H distribution follows a power law (Eq.~\ref{mag_dist}). Slopes of 0.46-0.54 are consistent with the H distributions for main belt asteroids. A slope of 0.7 was used in \cite{Cambioni2018} for similar extrapolations to compare their results to \cite{Whiteley1998}. The dotted line represents the point used for extrapolation, H=19.7 (see Fig.~\ref{fig:upper_lim}). The grey region depicts where N(\textless H) \textless\ 1, meaning the upper limit is essentially 0 ETs. Note, $\alpha$=0.46 gives the most conservative upper limits of the three power laws at the bright end, while $\alpha$=0.7 is most conservative at the faint end.}
  \label{fig:extrap_upper}
\end{figure}

\subsection{Upper Limit on ET Population}
\label{upperlimit}

\par To calculate an upper limit on the ET population, we injected the Strict set of synthetic ET objects (see Sec.~\ref{fakes}) following the methods described in Sec.~\ref{inject}. Of course, only sources that fell within the survey area could actually be injected into our images; 886 of the 20,555 objects were injected into our images (see Fig.~\ref{fig:inject_pos}). This means that our survey covered 4.3\% of this population.  However, our survey only covered the L5 point, meaning we have no direct constraints on the L4 population. Therefore, we calculated our upper limits for the L5 population only. Since there were 3726 L5 ETs in the full synthetic set, our survey coverage of L5 ETs specifically is 24.0\%.  However, assuming that the L4 and L5 points are symmetric, our upper limit is applicable to the L4 population as well. 

\par We then ran these images with synthetic ETs through our pipeline (Sec.~\ref{pipeline}) and counted the number of injected ETs that were then detected, linked, and flagged as an ET candidate. Fig.~\ref{fig:recovery} shows the recovery rate as a function of H. These plots are representative of the efficiency of our pipeline and our detection limit. Of note, there were 7 injected ET that were linked by the pipeline but were \textit{not} flagged as ETs due to to having speeds inconsistent with our 1 deg./day assumption. These objects are a small fraction of the nearly 1000 synthetic ETs injected into our images and were expected for synthetic objects with a small geocentric range (see Fig.~\ref{fig:et_speed}). These objects are not included in our recovery rate, as we would not have considered them to be an ET candidate.

\par Overall, we recovered about 50\% of the injected ETs. As expected, our recovery rate is much higher for the brighter objects and becomes very low at the faint end (12.8\% at H=20.4). In Fig. \ref{fig:recovery}, our limiting absolute magnitude appears to be H$\sim$20; however, H does not correspond directly to the V-band magnitudes of the injected ETs. 
\par From these recovery rates we can calculate an upper limit on the ET population adopting a Frequentist approach. Since we did not detect any ET candidates in our survey, by Poisson statistics, detecting 3 objects is within 3$\sigma$ of our result. Therefore we calculate our limit as 
\begin{equation}
\label{u_H}
    U(H) = \frac{3}{R(H) * C}
\end{equation}
where U(H) is our calculated upper limit, R(H) is the recovery rate as a function of H (see Fig. \ref{fig:recovery}) and C is the coverage of the L5 cloud for our survey (see Fig. \ref{fig:inject_pos}). Again, R(H) accounts for the performance of our pipeline and our limiting magnitude while C accounts for the limited area of our survey.

\par Fig.~\ref{fig:upper_lim} shows our calculated upper limits as a function of H and of size (assuming an albedo of 0.15). We also plot previously measured/estimated upper limits on the ET population at H=20.5. There are two distinct regimes in this recovery rate: flat at the bright end and sharp growth at the faint end. The upper limit becomes significantly higher at the faint end due to our limiting magnitude. In other words, the recovery rate becomes so low for the faintest objects that we only have relatively poor constraints on the population. Regardless, our calculated upper limit for the faintest H bin, H = 20.4 ($\sim$300 m), is 97. The bright end is essentially flat because the recovery rate was nearly 100\% for these bins; since C is the same for all bins and the recovery rate nearly the same, the calculated upper limit is constant ($\sim$15). 

\par If we assume that the underlying H distribution of the ET population follows a power law of index similar to that of the reported distributions for NEOs and main belt asteroids, we can calculate even more stringent (but model-dependent) upper limits for magnitudes in the flat regime. However, the true ET population may not follow such a distribution. If the ET population is primarily comprised of temporarily captured asteroids or NEOs, this assumption is likely reasonable. However, if the ET population is more similar to the ancient asteroid belt, it could be much shallower \citep{TSIRVOULIS201814} or depend on its collisional history \citep{BOTTKE2005111}. There is also evidence for differences in albedo and size distributions amongst NEO sub-groups themselves \citep{Mainzer_2012}. With our current sparse observational constraints, it is not possible to take these complexities into account; we instead assume the population follows a single power law of index in the range 0.46-0.7. 

\par Following this assumption, we calculate a cumulative H distribution as a power law normalized to our calculated upper limit for the faintest H bin still in the flat regime (corresponding to the dotted line in Fig.~\ref{fig:upper_lim}). We assume $\alpha$ = 0.46 and 0.54 as before (see Sec.~\ref{inject}) and 0.7 for comparison to \cite{Cambioni2018} (see Sec.~\ref{compare_prev}). Fig.~\ref{fig:extrap_upper} upper limits calculated by extrapolation Eq.~\ref{mag_dist} assuming each of these power laws. It is clear that with this method we obtain much more stringent constraints on bright ETs than in Fig.~\ref{fig:upper_lim}; we even obtain an upper limit of \textless\ 1 ET at H $\lesssim$ 15.5, 16.3, 17.2 for $\alpha$=0.46, 0.54, 0.7 respectively. We also calculate an upper limit of $\sim60-85$ (depending on the power law assumed) ETs with H \textless\ 19.7 (d$\sim$390 m assuming an albedo of 0.15). Note that $\alpha$=0.46 gives the most conservative upper limits for bright/large objects, whereas $\alpha$=0.7 gives the most conservative limits at the faint end.

\section{Discussion} \label{discussion}

\subsection{Summary}
\label{discussion_summary}

\par In this work, we present the results of our search for L5 ETs using the VR filter on DECam. In total, our survey covered 24 sq.~deg.~near the Earth-Sun L5 point. Our survey only consisted of $\sim$ 1 hour of observations, so the arcs for any observed Solar System objects were not long enough to fit an orbit or dynamically classify these objects. However, ETs are expected to have a distinctive rate of motion on the sky of 1 deg./day (Fig.~\ref{fig:et_speed}). Therefore, we developed the \textit{Bullseye} code to link together transient objects in our exposures and flag any tracklets with a similar speed as an ET candidate. 

\par With this survey, we found 27 tracklets that do not correspond to any previously known Solar System objects (see Table \ref{table:new_objects}). However, none of these objects were flagged as ET candidates. Based on this non-detection, we place upper limits (see Figs.~\ref{fig:upper_lim}, \ref{fig:extrap_upper}) on the ET population of $\sim60-85$ ETs with H\textless19.7 (corresponding to a size of $\sim$ 390 m assuming an albedo of 0.15) and 97 ETs with H=20.4 ($\sim$290 m).

\begin{figure*}
\includegraphics[width=\textwidth,keepaspectratio=true]{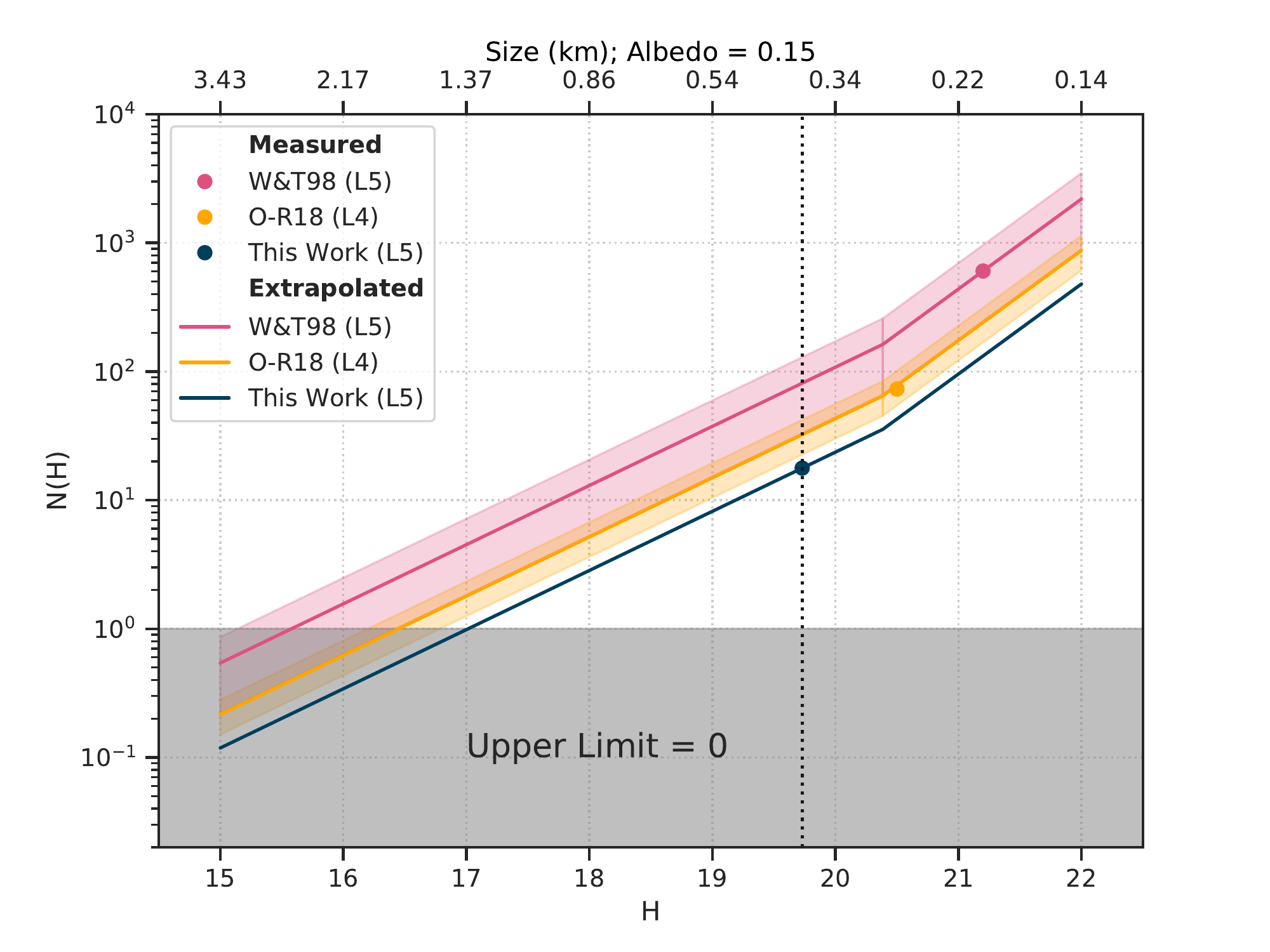}
  \caption[t]{Extrapolated upper limits (lines) based on the measured upper limit (circular points) for each survey. The results for this work are in blue, \cite{Cambioni2018} in yellow, and the upper limit calculated by \cite{Cambioni2018} based on the results in \cite{Whiteley1998} are in pink. The dotted line represents the point used for extrapolation for the results in this work, H=19.7. The grey region depicts where N(H) \textless 1, meaning the upper limit is essentially 0 ETs. The slope of the power law we use is $\alpha$=0.46 for H\textless20.39 and $\alpha$=0.7 for H\textgreater20.39, which gave the most conservative upper limits in Fig.~\ref{fig:extrap_upper}. Under these assumptions, the results from this work give the most stringent upper limit on the ET population to date (18 ETs with H=19.7 as compared to 32 and 81). For steeper power laws, the upper limits for this work are essentially consistent with O-R18.}
  \label{fig:extract_compare}
\end{figure*}

\subsection{Comparison to Previous Surveys}
\label{compare_prev}

\par There have been three previous dedicated searches for ETs \citep{Whiteley1998,Cambioni2018,hayabusa}. No ETs were found and no upper limits have been published in the Hayabusa2 search, so we can only compare to W\&T98 and O-R18. W\&T98 was similar to our search in that it was ground-based and covered L5, but it covered a much smaller area ($\sim$0.35 sq.~deg.). The O-R18 search was quite different in that is was conducted as the probe flew through the Lagrange point and it covered L4 rather than L5. However, we are still able to make comparisons to their results as we would expect the populations at the two Lagrange points to be symmetric. 

\par We first compared our faintest upper limit (at H=20.4) to the upper limits for W\&T98 and O-R18 as calculated in \cite{Cambioni2018} (see Fig.~\ref{fig:upper_lim}). Our upper limit of 97 ETs is slightly higher than the O-R18 result of 73 ETs at L4, but clearly more stringent than the upper limit calculated based on the W\&T98 search (194 ETs). 

\par However, in comparing to this single magnitude, we are comparing to our most poorly constrained upper limit. Therefore, we follow a process similar to that used in \cite{Cambioni2018} in which we extrapolate these other upper limits to a brighter H for which our survey has better constrained upper limits by assuming a power law distribution (see Sec. \ref{upperlimit}). This method requires assuming a slope parameter; we use  $\alpha$=0.46 for H\textless20.39 and $\alpha$=0.7 for H\textgreater20.39 as these values give the most conservative upper limits in Fig.~\ref{fig:extrap_upper}. 

\par Fig.~\ref{fig:extract_compare} shows the extrapolated upper limits using this model; for each survey, the circular point is the upper limit at the magnitude that we extrapolated from (coming from this work and \citet{Cambioni2018}). Again, our upper limits are clearly more stringent than the W\&T98 limits. Under these assumptions, our upper limits are also slightly more stringent than those from O-R18; if we used a power law with $\alpha$=0.7 only instead, our results are essentially consistent with the O-R18 results. At H=19.7, the point of extrapolation used for this work, we calculate an upper limit of 18, 32, and 81 ETs for this work, O-R18, and W\&T98 respectively. In short, although our survey does not have as faint of a limiting magnitude as the previous surveys, we are able to place more stringent limits on the L5 population at brighter magnitudes than \cite{Whiteley1998} and limits that are consistent to slightly more stringent than the OSIRIS-REx L4 search \citep{Cambioni2018}. 

\subsection{Missing Primordial ETs?}
\label{where_ets}

\par After this work, the existence/characterization of primordial ETs is still an open question. Even though it is almost certainly not primordial, \tk\ remains the only known ET. All of the current upper limits on the ET population, including from this work, suggest that there could still be tens to hundreds (depending on size) of ETs outside of the survey area of four independent searches. Deeper and broader coverage of both Lagrange clouds is necessary to say definitively whether or not other ETs exist. 

\par However, a dearth of additional ETs would not be entirely unexpected, especially for very faint/small objects. Numerical simulations have suggested that primordial ETs with H \textgreater 20.5 (assuming an albedo of 0.1) are likely to be destabilized due to the Yarkovsky effect within 4.5Gyr \citep{Zhou2018}. These results combined with previous non-detections of ETs lead \cite{Zhou2018} to conclude that it is very unlikely that primordial ETs of absolute magnitude H <= 20.5 (or their equivalent size limit) exist at the Lagrange points today, which could explain the current paucity of ETs. This work, as well as previous surveys, are sensitive to objects primarily above this limit; thus, deeper searches would allow us to test this theoretical limit. 

\par It is important to note that it is also possible for asteroids to be temporarily trapped in an ET-like orbit, like \tk. \cite{MORAIS20021} found that such objects are likely quite rare though; through their simulations they calculated an upper limit of 0.65$\pm$0.12 and 16.3$\pm$3.0 captured Earth co-orbitals for H\textless18 and H\textless22 respectively. Based on these results, it seems possible that \tk\ truly is the only Earth co-orbital. However, it is also possible that the Yarkovsky effect (not included in those simulations) could in fact improve the capture efficiency for transient co-orbital ETs \citep{Malhotra2019}.  

\par Ultimately, in the absence of stronger observational constraints, there is still little we can definitely say about primordial ETs. Broader and deeper searches are still needed in order to further constrain this population.  




\section*{Acknowledgements}

This material is based upon work supported by the National Aeronautics and Space
Administration under Grant No. NNX17AF21G issued through the SSO Planetary Astronomy Program. L.M., J.C.B, and S.J.H are supported by the NSF Graduate Research Fellowship Grant No. DGE 1256260. R.M.~acknowledges support from NASA Nexus for Exoplanet System Science (NExSS; grant NNX15AD94G).

Based in part on observations at Cerro Tololo InterAmerican Observatory, NSF’s National Optical-Infrared Astronomy Research Laboratory, which is operated by the Association of Universities for Research in Astronomy (AURA) under a cooperative agreement with the National Science Foundation. These observations were taken under NOAO 2018A-0177, Alex Parker et al.

We thank the referee for helpful and constructive comments, which improved this paper. We would like to thank Hsing Wen Lin and Tali Khain for their useful comments on this paper.




\bibliographystyle{mnras}
\bibliography{mnras_template} 




\appendix

\section{Survey Pointings}
\begin{table}
\caption{First Half Survey Pointings. Columns: (1) Name of the Field. There were 8 distinct fields in this survey, observed 8 times each; (2) RA of exposure center (hours); (3) Dec of exposure center (degrees); (4) Start time of exposure (UTC). Every exposure was taken on 2018-06-16.}

\begin{center}
\begin{tabular}{llllr}
\hline
Field & RA & Dec & Time \\
\hline
ET\_0 &  148.2684 &  13.71537 &  22:45:35 \\
 ET\_0 &  148.2686 &  13.71430 &  22:46:18 \\
 ET\_1 &  146.8672 &  12.61325 &  22:47:02 \\
 ET\_1 &  146.8663 &  12.61278 &  22:47:45 \\
 ET\_2 &  148.6678 &  11.91204 &  22:48:28 \\
 ET\_2 &  148.6661 &  11.91206 &  22:49:11 \\
 ET\_3 &  150.0658 &  13.21204 &  22:49:54 \\
 ET\_3 &  150.0649 &  13.21155 &  22:50:36 \\
 ET\_4 &  151.8663 &  12.61203 &  22:51:19 \\
 ET\_4 &  151.8647 &  12.61239 &  22:52:03 \\
 ET\_5 &  150.4654 &  11.31283 &  22:52:46 \\
 ET\_5 &  150.4657 &  11.31343 &  22:53:29 \\
 ET\_6 &  152.2671 &  10.71340 &  22:54:12 \\
 ET\_6 &  152.2659 &  10.71455 &  22:54:56 \\
 ET\_7 &  153.6655 &  12.01450 &  22:55:38 \\
 ET\_7 &  153.6665 &  12.01432 &  22:56:20 \\
 ET\_0 &  148.2658 &  13.71357 &  22:57:16 \\
 ET\_0 &  148.2666 &  13.71279 &  22:58:00 \\
 ET\_1 &  146.8659 &  12.61219 &  22:58:46 \\
 ET\_1 &  146.8652 &  12.61255 &  22:59:29 \\
 ET\_2 &  148.6659 &  11.91192 &  23:00:11 \\
 ET\_2 &  148.6661 &  11.91251 &  23:00:53 \\
 ET\_3 &  150.0671 &  13.21225 &  23:01:40 \\
 ET\_3 &  150.0665 &  13.21313 &  23:02:23 \\
 ET\_4 &  151.8678 &  12.61350 &  23:03:06 \\
 ET\_4 &  151.8686 &  12.61258 &  23:03:50 \\
 ET\_5 &  150.4693 &  11.31235 &  23:04:32 \\
 ET\_5 &  150.4683 &  11.31208 &  23:05:15 \\
 ET\_6 &  152.2692 &  10.71229 &  23:05:57 \\
 ET\_6 &  152.2691 &  10.71294 &  23:06:40 \\
 ET\_7 &  153.6702 &  12.01269 &  23:07:22 \\
 ET\_7 &  153.6697 &  12.01314 &  23:08:04 \\
  \end{tabular}
 \end{center}
 \label{table:pointings}
\end{table}

\begin{table}
\caption{Second Half Survey Pointings. Columns: (1) Name of the Field. There were 8 distinct fields in this survey, observed 8 times each; (2) RA of exposure center (hours); (3) Dec of exposure center (degrees); (4) Start time of exposure (UTC). Every exposure was taken on 2018-06-16.}
\begin{center}
\begin{tabular}{llllr}
\hline
Field & RA & Dec & Time \\
\hline
 ET\_0 &  148.2691 &  13.71246 &  23:08:59 \\
 ET\_0 &  148.2682 &  13.71292 &  23:09:41 \\
 ET\_1 &  146.8680 &  12.61210 &  23:10:24 \\
 ET\_1 &  146.8683 &  12.61332 &  23:11:07 \\
 ET\_2 &  148.6696 &  11.91296 &  23:11:49 \\
 ET\_2 &  148.6698 &  11.91222 &  23:12:33 \\
 ET\_3 &  150.0690 &  13.21317 &  23:13:16 \\
 ET\_3 &  150.0684 &  13.21276 &  23:13:58 \\
 ET\_4 &  151.8679 &  12.61339 &  23:14:41 \\
 ET\_4 &  151.8673 &  12.61308 &  23:15:23 \\
 ET\_5 &  150.4674 &  11.31342 &  23:16:05 \\
 ET\_5 &  150.4680 &  11.31393 &  23:16:47 \\
 ET\_6 &  152.2689 &  10.71375 &  23:17:29 \\
 ET\_6 &  152.2697 &  10.71322 &  23:18:12 \\
 ET\_7 &  153.6695 &  12.01254 &  23:18:53 \\
 ET\_7 &  153.6707 &  12.01305 &  23:19:37 \\
 ET\_0 &  148.2691 &  13.71227 &  23:20:34 \\
 ET\_0 &  148.2684 &  13.71222 &  23:21:17 \\
 ET\_1 &  146.8683 &  12.61170 &  23:21:59 \\
 ET\_1 &  146.8674 &  12.61258 &  23:22:43 \\
 ET\_2 &  148.6676 &  11.91242 &  23:23:25 \\
 ET\_2 &  148.6669 &  11.91336 &  23:24:08 \\
 ET\_3 &  150.0660 &  13.21263 &  23:24:52 \\
 ET\_3 &  150.0669 &  13.21335 &  23:25:35 \\
 ET\_4 &  151.8677 &  12.61336 &  23:26:17 \\
 ET\_4 &  151.8668 &  12.61394 &  23:27:01 \\
 ET\_5 &  150.4666 &  11.31324 &  23:27:43 \\
 ET\_5 &  150.4675 &  11.31401 &  23:28:26 \\
 ET\_6 &  152.2673 &  10.71314 &  23:29:09 \\
 ET\_6 &  152.2673 &  10.71354 &  23:29:51 \\
 ET\_7 &  153.6683 &  12.01315 &  23:30:34 \\
 ET\_7 &  153.6688 &  12.01308 &  23:31:16 \\
\end{tabular}
\end{center}

\end{table}

\section{Calculating Upper Limits Under Different Assumptions of the ET Population}
In Sec. \ref{upperlimit}, we calculated various upper limits on the ET population, which required making several assumption about the population, namely its physical extent and size distribution. Under different assumptions the calculated upper limits could deviate from what was presented here. Therefore, we aim to provide enough information such that the reader could calculate upper limits based on our observations but under their own assumptions. 

\par We calculated upper limits from our recovery rates using Eq. \ref{u_H}. This equation depends on our calculated recovery rates and our assumed survey coverage of the L5 cloud. We determined this coverage to be 24\% based on our synthetic population (see Sec. \ref{fakes}). However, these upper limits may be calculated with a different coverage value but the same measured values of the recovery rates.  These values appear in Fig. \ref{fig:recovery}, but we include them explicitly here in \ref{table:recovery}.

We also assume that the ET H/size distribution follows a power law (Eq.~\ref{mag_dist}) in order to calculate even more stringent upper limits on the ET population, particularly for bright/large objects. We performed the extrapolation using the non-cumulative power law distribution normalized to the upper limit we calculated for our point of extrapolation: 
\begin{equation}
\label{mag_dist_non_cum}
    N(H) = \frac{U(H_e)}{10 ^ {\alpha H_e}} * 10 ^ {\alpha H}
\end{equation}
where H is the absolute magnitude, $H_{e}$ is the extrapolation point, $U(H_{e})$ is the upper limit calculated at that extrapolation point, and $\alpha$ is a slope parameter. In Sec. \ref{upperlimit}, we use $H_{e}$ = 19.7, $U(H_{e})$ = 18, and $\alpha$ = 0.46 for H < 20.39 and $\alpha$ = 0.7 for H > 20.39. However, this method may still be used for different assumptions of slope values, extrapolation points, or other functional forms of the ET H distribution given our calculated upper limit at the H bin used for extrapolation; these values appear in Fig. \ref{fig:upper_lim}, but we include them explicitly here in \ref{table:recovery}. 
\begin{table}
\caption{Values needed to calculate ET upper limits. Columns: (1) H bin; (2) Recovery rate of injected Earth Trojans (see Sec. \ref{inject}, \ref{upperlimit}) such that 1 = 100\% of objects were recovered. If this column contains a dashed line, there were no injected objects in that H bin.; (3) Calculated upper limit without extrapolation for the number of ETs in each H bin (see Sec. \ref{upperlimit}). If this column contains a dashed line, there were no injected objects in that H bin, so we could not calculate an upper limit.}

\begin{center}
\begin{tabular}{ccc}
\hline
H & R(H) & U(H) \\
\hline
15.11 & 0.40 & 31.54 \\
15.33 & - - - & - - - \\
15.55 & 1.00 & 12.62 \\
15.77 & - - - & - - -\\
15.99 & 0.33 & 37.85 \\
16.21 & - - - & - - -\\
16.43 & 1.00 & 12.62 \\
16.65 & 0.50 & 25.23 \\
16.87 & 0.67 & 18.92 \\
17.09 & 0.83 & 15.14 \\
17.31 & 0.89 & 14.19 \\
17.53 & 0.67 & 18.92 \\
17.75 & 0.80 & 15.77 \\
17.97 & 0.90 & 14.02 \\
18.19 & 0.94 & 13.36 \\
18.41 & 0.75 & 16.82 \\
18.63 & 0.91 & 13.92 \\
18.85 & 0.77 & 16.33 \\
19.07 & 0.84 & 14.94 \\
19.29 & 0.62 & 20.19 \\
19.51 & 0.76 & 16.68 \\
19.73 & 0.71 & 17.73 \\
19.95 & 0.53 & 24.01 \\
20.17 & 0.31 & 40.27 \\
20.39 & 0.13 & 97.15 \\
  \end{tabular}
 \end{center}
 \label{table:recovery}
\end{table}



\bsp	
\label{lastpage}
\end{document}